\documentclass[a4paper,accepted=2020-10-05]{quantumarticle}

\pdfoutput=1

\usepackage{xkeyval, etoolbox, geometry, xcolor, fancyhdr, tikz, hyperref, ltxgrid,ltxcmds}
\usepackage{preamble}
\usepackage[numbers,sort&compress]{natbib}

\begin{document}

\title{A quantum extension of SVM-perf for training nonlinear SVMs in almost linear time}

\author{Jonathan Allcock}
\email{jonallcock@tencent.com}
\affiliation{Tencent Quantum Laboratory}
\author{Chang-Yu Hsieh}
\email{kimhsieh@tencent.com}
\affiliation{Tencent Quantum Laboratory}

\date{October 2020}

\begin{abstract}
  We propose a quantum algorithm for training nonlinear support vector machines (SVM) for feature space learning where classical input data is encoded in the amplitudes of quantum states. Based on the classical SVM-perf algorithm of Joachims \cite{joachims2006training}, our algorithm has a running time which scales linearly in the number of training examples $m$ (up to polylogarithmic factors) and applies to the standard soft-margin $\ell_1$-SVM model. In contrast, while classical SVM-perf has demonstrated impressive performance on both linear and nonlinear SVMs, its efficiency is guaranteed only in certain cases: it achieves linear $m$ scaling only for linear SVMs, where classification is performed in the original input data space, or for the special cases of low-rank or shift-invariant kernels.  Similarly, previously proposed quantum algorithms either have super-linear scaling in $m$, or else apply to different SVM models such as the hard-margin or least squares $\ell_2$-SVM which lack certain desirable properties of the soft-margin $\ell_1$-SVM model. We classically simulate our algorithm and give evidence that it can perform well in practice, and not only for asymptotically large data sets. 
\end{abstract}

\maketitle

\section{Introduction}

Support vector machines (SVMs) are powerful supervised learning models which perform classification by identifying a decision surface which separates data according to their labels \cite{boser1992training,cortes1995support}. While classifiers based on deep neural networks have increased in popularity in recent years, SVM-based classifiers maintain a number of advantages which make them an appealing choice in certain situations. SVMs are simple models with a smaller number of trainable parameters than neural networks, and thus can be less prone to overfitting and easier to interpret.  Furthermore, neural network training may often get stuck in local minima, whereas SVM training is guaranteed to find a global optimum \cite{burges1998tutorial}.   For problems such as text classification which involve high dimensional but sparse data, linear SVMs --- which seek a separating hyperplane in the same space as the input data --- have been shown to perform extremely well, and training algorithms exist which scale efficiently, i.e. linearly in \cite{ferris2002interior, mangasarian2001lagrangian, keerthi2005modified}, or even independent of \cite{shalev2011pegasos}, the number of training examples $m$. 

In more complex cases, where a nonlinear decision surface is required to classify the data successfully,  nonlinear SVMs can be used, which seek a separating hyperplane in a higher dimensional feature space.  Such feature space learning typically makes use of the kernel trick \cite{scholkopf2001learning},  a method enabling inner product computations in high or even infinite dimensional spaces to be performed implicitly, without requiring the explicit and resource intensive computation of the feature vectors themselves.  

While powerful, the kernel trick comes at a cost: many classical algorithms based on this method scale poorly with $m$.  Indeed, storing the full kernel matrix $K$ in memory itself requires $O(m^2)$ resources, making subquadratic training times impossible by brute-force computation of $K$. When $K$ admits a low-rank approximation though, sampling-based approaches such as the Nystrom method \cite{williams2001using} or incomplete Cholesky factorization \cite{fine2001efficient} can be used to obtain $O(m)$ running times, although it may not be clear a priori whether such a low-rank approximation is possible. Another special case corresponds to so-called shift-invariant kernels \cite{rahimi2008random}, which include the popular Gaussian radial basis function (RBF) kernel, where classical sampling techniques can be used to map the high dimensional data into a random low dimensional feature space, which can then be trained by fast linear methods. This method has empirically competed favorably with more sophisticated kernel machines in terms of classification accuracy, at a fraction of the training time. While such a method seems to strike a balance between linear and nonlinear approaches, it cannot be applied to more general kernels.  In practice, advanced solvers employ multiple heuristics to improve their performance, which makes rigorous analyses of their performance difficult. However, methods like SVM-Light \cite{joachims1999advances}, SMO \cite{platt1999smo}, LIBSVM \cite{chang2011libsvm} and SVMTorch \cite{collobert2001svmtorch} still empirically scale approximately quadratically with $m$ for nonlinear SVMs. 

The state-of-the-art in terms of provable computational complexity is the Pegasos algorithm \cite{shalev2011pegasos}. Based on stochastic sub-gradient descent, Pegasos has constant running time for linear SVMs. For nonlinear SVMs, Pegasos has $O(m)$ running time, and is not restricted to low-rank or shift-invariant kernels. However, while experiments show that Pegasos does indeed display outstanding performance for linear SVMs, for nonlinear SVMs it is outperformed by other benchmark methods on a number of datasets. On the other hand, the SVM-perf algorithm of Joachims \cite{joachims2006training} has been shown to outperform similar benchmarks \cite{joachims2009sparse}, although it does have a number of theoretical drawbacks compared with Pegasos.  SVM-perf has $O(m)$ scaling for linear SVMs, but an efficiency for nonlinear SVMs which either depends on heuristics, or on the presence of a low-rank or shift-invariant kernel, where linear in $m$ scaling can also be achieved. However, given the strong empirical performance of SVM-perf, it serves as a strong starting point for further improvements, with the aim of overcoming the restrictions in its application to nonlinear SVMs.

Can quantum computers implement SVMs more effectively than classical computers? Rebentrost and Lloyd were the first to consider this question \cite{rebentrost2014quantum}, and since then numerous other proposals have been put forward \cite{havlivcek2019supervised, schuld2019quantum, kerenidis2019quantum, arodz2019quantum, li2019sublinear}. While the details vary, at a high level these quantum algorithms aim to bring benefits in two main areas: i) faster training and evaluation time of SVMs or ii) greater representational power by encoding the high dimensional feature vectors in the amplitudes of quantum states. Such quantum feature maps enable high dimensional inner products to be computed directly and, by sidestepping the kernel trick, allow classically intractable kernels to be computed. These proposals are certainly intriguing, and open up new possibilities for supervised learning. However, the proposals to date with improved running time dependence on $m$ for nonlinear SVMs do not apply to the standard soft-margin $\ell_1$-SVM model, but rather to variations such as least squares $\ell_2$-SVMs\cite{rebentrost2014quantum} or hard-margin SVMs \cite{li2019sublinear}.  While these other models are useful in certain scenarios, soft-margin $\ell_1$-SVMs have two properties - sparsity of weights and robustness to noise - that make them preferable in many circumstances.

In this work we present a method to extend SVM-perf to train nonlinear soft-margin $\ell_1$-SVMs with quantum feature maps in a time that scales linearly (up to polylogarithmic factors) in the number of training examples, and which is not restricted to low-rank or shift-invariant kernels.  
Provided that one has quantum access to the classical data, i.e. quantum random access memory (qRAM) \cite{giovannetti2008quantum, prakash2014quantum}, quantum states corresponding to sums of feature vectors can be efficiently created, and then standard methods employed to approximate the inner products between such quantum states.  As the output of the quantum procedure is only an approximation to a desired positive semi-definite (p.s.d.) matrix, it is not itself guaranteed to be p.s.d., and hence an additional classical projection step must be carried out to map on to the p.s.d. cone at each iteration.

Before stating our result in more detail, let us make one remark. It has recently been shown by Tang \cite{tang2019quantum} that the data-structure required for efficient qRAM-based inner product estimation would also enable such inner products to be estimated classically, with only a polynomial slow-down relative to quantum, and her method has been employed to de-quantize a number of quantum machine learning algorithms \cite{tang2019quantum, tang2018quantum02, gilyen2018quantum}  based on such data-structures. However, in practice, polynomial factors can make a difference, and an analysis of a number of such quantum-inspired classical algorithms \cite{arrazola2019quantum}  concludes that care is needed when assessing their performance relative to the quantum algorithms from which they were inspired.  More importantly, in this current work, the quantum states produced using qRAM access are subsequently mapped onto a larger Hilbert space before their inner products are evaluated.  This means that the procedure cannot be de-quantized in the same way.

\section{Background and Results}

Let $S = \{ (\mathbf{x}_1, y_1), \ldots (\mathbf{x}_m, y_m)\}$ be a data set with $\mathbf{x}_i\in\mb{R}^d$, and labels $y_i\in\{+1,-1\}$. Let $\Phi:\mb{R}^d\ra \mathcal{H}$ be a feature map where $\+H$ is a real Hilbert space (of finite or infinite dimension) with inner product $\ip{\cdot\,}{\cdot}$, and let $K:\mb{R}^{d}\times \mb{R}^d \ra \mb{R}$ be the associated kernel function defined by $K(\.x,\.y)\defeq \langle \Phi(\.x), \Phi(\.y)\rangle$. Let $R = \max_i\norm{\Phi(\.x_i)}$ denote the largest $\ell_2$ norm of the feature mapped vectors. In what follows, $\norm{\cdot}$ will always refer to the $\ell_2$ norm, and other norms will be explicitly differentiated.

\subsection{Support Vector Machine Training}

Training a soft-margin $\ell_1$-SVM with parameter $C > 0$ corresponds to solving the following optimization problem:

\begin{op}{(SVM Primal)}\label{op:svm-p}
\eq{
    \quad \underset{\.w\in \mathcal H,\ \xi_i\ge 0}{\min} &\quad  \frac{1}{2}\ip{\.w}{\.w} + \frac{C}{m}\sum_{i=1}^m \xi_i \\
    s.t. &\quad  y_i \ip{\.w}{\Phi(x_i)} \ge 1- \xi_i \quad \forall i=1,\ldots, m
    }
\end{op}
Note that, following \cite{joachims2006training}, we divide $\sum_i\xi_i$ by $m$ to capture how $C$ scales with the training set size. The trivial case $\Phi(\.x) = \.x$ corresponds to a linear SVM, i.e. a separating hyperplane is sought in the original input space.   When one considers feature maps $\Phi(\.x)$ in a high dimensional space, it is more practical to consider the dual optimization problem, which is expressed in terms of inner products, and hence the kernel trick can be employed.

\begin{op}{(SVM Dual)}\label{op:svm-d}
\eq{
    \underset{\mathbf{\alpha}}{\max} &\quad  -\frac{1}{2}\sum_{i,j=1}^m y_i\alpha_i y_j\alpha_j K(\.x_i,\.x_j) + \sum_{i=1}^m \alpha_i \\ 
    \text{s.t.} &\quad 0 \le \alpha_i \le \frac{C}{m}\qquad \forall i=1,\ldots, m
}
\end{op}
This is a convex quadratic program with box constraints, for which many classical solvers are available, and which requires time polynomial in $m$ to solve. For instance, using the barrier method \cite{boyd2004convex} a solution can be found to within $\eps_b$ in time $O(m^4\log(m/\eps_b))$.  Indeed, even the computation of the kernel matrix $K$ takes time $\Theta(m^2)$, so obtaining subquadratic training times via direct evaluation of $K$ is not possible. 

\subsection{Structural SVMs}
Joachims \cite{joachims2006training} showed that an efficient approximation algorithm - with running time $O(m)$ - for linear SVMs could be obtained by considering a slightly different but related model known as a \textit{structural SVM} \cite{tsochantaridis2005large}, which makes use of linear combinations of label-weighted feature vectors:

\begin{definitionenv}\label{def:psi-c} For a given data set $S = \left\{(\.x_1,y_1),\ldots, (\.x_m,y_m)\right\}$, feature map $\Phi$, and $\.c\in\{0,1\}^m$, define
\eq{
\Psi_{\.c}\defeq \frac{1}{m}\sum_{i=1}^m c_i y_i \Phi(\.x_i)
}
\end{definitionenv}
With this notation, the structural SVM primal and dual optimization problems are:
\begin{op}{(Structural SVM Primal)}\label{op:struct-svm-p}
\eq{
    \underset{\.w\in \mathcal H,\ \xi\ge 0}{\min} & \quad  P(\.w, \xi) \defeq \frac{1}{2}\ip{\.w}{\.w} + C\xi \\
    \text{s.t.} &\quad  \frac{1}{m}\sum_{i=1}^m c_i - \ip{\.w}{\Psi_{\.c}} \le \xi,\qquad \forall \.c\in\{0,1\}^m
    }
\end{op}

\begin{op}{(Structural SVM Dual)}\label{op:struct-svm-d}
\eq{
    \underset{\alpha \ge 0}{\max} &\,\,  D(\alpha) \defeq -\frac{1}{2}\sum_{\.c,\.c'\in\{0,1\}^m} \alpha_{\.c} \alpha_{\.c'}J_{\.c\.c'} + \sum_{\.c\in\{0,1\}^m} \frac{\norm{\.c}_1}{m}\alpha_{\.c} \\ 
    \text{s.t.} &\sum_{\.c\in\{0,1\}^m}\alpha_{\.c} \le C
}
where $J_{\.c\.c'} = \ip{\Psi_{\.c}}{\Psi_{\.c'}}$ and $\norm{\cdot}_1$ denotes the $\ell_1$-norm.
\end{op}
 Whereas the original SVM problem OP \ref{op:svm-p} is defined by $m$ constraints and $m$ slack variables $\xi_i$,  the structural SVM  OP \ref{op:struct-svm-p} has only one slack variable $\xi$ but $2^m$ constraints, corresponding to each possible binary vector $\.c\in\{0,1\}^m$.  In spite of these differences, the solutions to the two problems are equivalent in the following sense.

\begin{restatable}[Joachims \cite{joachims2006training}]{theorem}{theoremjoachim01}
\label{thm:joachim}
Let $(\.w^*,\xi^*_1,\ldots, \xi^*_m)$ be an optimal solution of OP \ref{op:svm-p}, and let $\xi^* = \frac{1}{m}\sum_{i=1}^m\xi^*_i$. Then $(\.w^*,\xi^*)$ is an optimal solution of OP \ref{op:struct-svm-p} with the same objective function value. Conversely, for any optimal solution $(\.w^*,\xi^*)$ of OP \ref{op:struct-svm-p}, there is an optimal solution $(\.w^*,\xi^*_1,\ldots, \xi^*_m)$ of OP \ref{op:svm-p} satisfying $\xi^* = \frac{1}{m}\sum_{i=1}^m\xi_i$, with the same objective function value.
\end{restatable}

While elegant, Joachims' algorithm can achieve $O(m)$ scaling only for linear SVMs --- as it requires explicitly  computing a set of vectors $\{\Psi_{\.c}\}$ and their inner products --- or to shift-invariant or low-rank kernels where sampling methods can be employed. For high dimensional feature maps $\Phi$ not corresponding to shift invariant kernels, computing $\Psi_{\.c}$ classically is inefficient.  We propose instead to embed the feature mapped vectors $\Phi(\.x)$ and linear combinations $\Psi_{\.c}$ in the amplitudes of quantum states, and compute the required inner products efficiently using a quantum computer. 

\subsection{Our Results}

In Section \ref{sec:methods} we will formally introduce the concept of a quantum feature map. For now it is sufficient to view this as a quantum circuit which, in time $T_{\Phi}$, realizes a feature map $\Phi:\mb{R}^d\ra\+H$, with maximum norm $\max_{\.x}\norm{\Phi(\.x)} = R$, by mapping the classical data into the state of a multi-qubit system.

Our first main result is a quantum algorithm with running time linear in $m$ that generates an approximately optimal solution for the structural SVM problem. By Theorem \ref{thm:joachim}, this is equivalent to solving the original soft-margin $\ell_1$-SVM.  

\vspace{0.2cm}
\noindent {\em \textbf{Quantum nonlinear SVM training}: [See Theorems \ref{thm:hybrid-algo-iterations} and \ref{thm:hybrid-algo-time}] There is a quantum algorithm that, with probability at least $1-\delta$,  outputs $\hat{\alpha}$ and $\hat{\xi}$ such that if $(\.w^*, \xi^*)$ is the optimal solution of OP \ref{op:struct-svm-p}, then 
\eq{
P(\hat{\.w}, \hat{\xi}) - P(\.w^*,\xi^*) \le \min\left\{\frac{C\epsilon}{2},\frac{\epsilon^2}{8R^2}\right\}
}
where $\hat{\.w}\defeq \sum_{\.c} \hat{\alpha}_{\.c} \Psi_{\.c}$, and  $(\hat{\.w}, \hat{\xi} + 3\epsilon)$ is feasible for OP \ref{op:struct-svm-p}. The running time is
\eq{
 \tilde{O}\lp \frac{CR^3\log(1/\delta)}{\Psi_{\min}} \lp   \frac{t^2_{\max}}{\epsilon}\cdot m + t_{\max}^5 \rp T_\Phi  \rp 
}
where $t_{\max} = \max\left\{ \frac{4}{\epsilon},\frac{16CR^2}{\epsilon^2} \right\}$, $T_\Phi$ is the time required to compute feature map $\Phi$ on a quantum computer and $\Psi_{\min}$ is a term that depends on both the data as well as the choice of quantum feature map.}

Here and in what follows, the tilde big-O notation hides polylogarithmic terms. In the Simulation section we show that, in practice, the running time of the algorithm can be significantly faster than the theoretical upper-bound. The solution $\hat{\alpha}$ is a $t_{\max}$-sparse vector of total dimension $2^{m}$. Once it has been found, a new data point $\.x$ can be classified according to
\eq{
y_{pred} &= \sgn \ip{\sum_{\.c}\hat{\alpha}_{\.c}\Psi_{\.c}}{ \Phi(\.x)} \\
&= \sgn \lp \sum_{\.c}\hat{\alpha}_{\.c}\sum_{i=1}^m \frac{c_iy_i}{m} \ip{\Phi(\.x_i)}{\Phi(\.x)}\rp
}
where $y_{pred}$ is the predicted label of $\.x$. This is a sum of $O(m t_{\max})$ inner products in feature space, which classical methods require time $O(m t_{\max})$ to evaluate in general.  Our second result is a quantum algorithm for carrying out this classification with running time independent of $m$.   

\vspace{0.2cm} 
\noindent{\em \textbf{Quantum nonlinear SVM classification}: [See Theorem \ref{thm:q-classification}] There is a quantum algorithm which, in time 
\eq{
\tilde{O}\lp\frac{CR^3\log(1/\delta)}{\Psi_{\min}}\frac{t_{\max}}{\epsilon}T_{\Phi}\rp
}
outputs, with probability at least $1-\delta$, an estimate to $ \ip{\sum_{\.c}\hat{\alpha}_{\.c}\Psi_{\.c}}{ \Phi(\.x)}$ to within $\epsilon$ accuracy.  
}
The sign of the output is then taken as the predicted label.


\section{Methods}\label{sec:methods}

Our results are based on three main components: Joachims' linear time classical algorithm SVM-perf, quantum feature maps, and efficient quantum methods for estimating inner products of linear combinations of high dimensional vectors.

\subsection{SVM-perf: a linear time algorithm for linear SVMs}

On the surface, the structural SVM problems OP \ref{op:struct-svm-p} and OP \ref{op:struct-svm-d} look more complicated to solve than the original SVM problems OP \ref{op:svm-p} and OP \ref{op:svm-d}.  However, it turns out that the solution $\alpha^*$ to OP \ref{op:struct-svm-d} is highly sparse and, consequently, the structural SVM admits an efficient algorithm. Joachims' original procedure is presented in Algorithm \ref{alg:struct-svm}.

\begin{Algorithm}
\captionsetup{labelsep=newline,format=hline_caption}
\caption{SVM-perf \cite{joachims2006training}:  Training structural SVMs via OP \ref{op:struct-svm-p}}
\label{alg:struct-svm}
\begin{algorithmic}
\STATE \textbf{Input: }Training set $S = \left\{ (\.x_1, y_1), \ldots, (\.x_m, y_m)\right\}$, SVM  hyperparameter $C> 0$, tolerance $\epsilon> 0$, $\.c\in\{0,1\}^m$. 
\STATE 
\STATE $\+W \gets \{\.c\}$
\REPEAT
\STATE  $(\.w,\xi)\gets \arg\min_{\.w,\xi\ge 0}\frac{1}{2}\ip{\.w}{\.w} + C\xi$
\STATE $\quad\quad\quad \text{s.t.}\quad \forall \.c\in\+W:\quad  \frac{1}{m}\sum_{i=1}^m c_i - \ip{\.w}{\Psi_{\.c}} \le \xi$

\FOR{$i=1,\ldots, m$}
    \STATE  $c^*_i \gets \begin{cases}1 & y_i\lp \.w^T \.x_i\rp  < 1\\ 0 & \text{otherwise}\end{cases}$
\ENDFOR
\STATE  $\+W\gets \+W\cup \{\.c^*\}$
\UNTIL{ $\frac{1}{m}\sum_{i=1}^m c^*_i - \sum \alpha_{\.c'\in\+W} \ip{\Psi_{\.c'}}{\Psi_{\.c^*}}  \le \xi + \epsilon$}
\STATE 
\STATE  \textbf{Output: }  $(\.w,\xi)$
\end{algorithmic}
\hrulefill
\end{Algorithm}

The main idea behind Algorithm \ref{alg:struct-svm} is to iteratively solve successively more constrained versions of problem OP \ref{op:struct-svm-p}. That is, a working set of indices $\+W\subseteq\{0,1\}^m$ is maintained such that, at each iteration, the solution $(\.w,\xi)$ is only required to satisfy the constraints $\frac{1}{m}\sum_{i=1}^m c_i - \ip{\.w}{\Psi_c} \le \xi$ for $\.c\in\+W$.  The inner \textbf{for} loop then finds a new index $\.c^*$ which corresponds to the maximally violated constraint in OP \ref{op:struct-svm-p}, and this index is added to the working set.  The algorithm proceeds until no constraint is violated by more than $\epsilon$.  It can be shown that each iteration must improve the value of the dual objective by a constant amount, from which it follows that the algorithm terminates in a number of rounds independent of $m$.  

\begin{restatable}[Joachims \cite{joachims2006training}]{theorem}{theoremjoachim02}
 Algorithm \ref{alg:struct-svm} terminates after at most $\max\left\{\frac{2}{\epsilon}, \frac{8CR^2}{\epsilon^2}\right\}$ iterations, where $R\defeq\max_i\norm{\.x_i}$ is the largest $\ell_2$-norm of the training set vectors. For any training set $S$ and any $\epsilon >0$, if $(\.w^*, \xi^*)$ is an optimal solution of OP \ref{op:struct-svm-p}, then Algorithm \ref{alg:struct-svm} returns a point $(\.w, \xi)$ that has a better objective value than $(\.w^*, \xi)$, and for which $(\.w,\xi + \epsilon)$ is feasible in OP \ref{op:struct-svm-p}.
\end{restatable}

In terms of time cost, each iteration $t$ of the algorithm involves solving the restricted optimization problem
\eq{
(\.w,\xi)&=\arg\min_{\.w,\xi\ge 0}\frac{1}{2}\ip{\.w}{\.w} + C\xi\\
\text{s.t.}&\quad \forall \.c\in\+W:\quad  \frac{1}{m}\sum_{i=1}^m c_i - \ip{\.w}{\Psi_c}\le \xi
}
which is done in practice by solving the corresponding dual problem, i.e. the same as OP \ref{op:struct-svm-d} but with summations over $\.c\in\+W$ instead of over all $\.c\in\{0,1\}$. This involves computing 
\begin{itemize}
    \item $O(t^2)$ matrix elements $J_{\.c\.c'} = \ip{\Psi_{\.c}}{\Psi_{\.c'}}$
    \item $m$ inner products $\ip{\.w}{\.x_i} = \ip{\sum_{\.c}\alpha_{\.c}\Psi_{\.c}}{\.x_i}$
\end{itemize}
where $\alpha_{\.c}$ is the solution to the dual of the optimization problem in the body of Algorithm \ref{alg:struct-svm}. In the case of linear SVMs, $\Psi_{\.c} = \frac{1}{m}\sum_{i=1}^m c_i y_i \.x_i$ and $\ip{\Psi_{\.c}}{\Psi_{\.c'}}$ can each be explicitly computed in time $O(m)$.  The cost of computing matrix $J$ is $O(mt^2)$, and subsequently solving the dual takes time polynomial in $t$. As Joachims showed that $t \le \max\left\{\frac{2}{\epsilon}, \frac{8C\max_i\norm{\.x_i}^2}{\epsilon^2}\right\}$, and since $\ip{\.w}{\.x_i}$ can be computed in time $O(d)$, the entire algorithm therefore has running time linear in $m$. Note that Joachims considered the special case of $s$-sparse data vectors, for which $\ip{\.w}{\.x_i}$ can be computed in time $O(s)$ rather than $O(d)$. In what follows we will not consider any sparsity restrictions.

For nonlinear SVMs, the feature maps $\Phi(\.x_i)$ may be of very large dimension, which precludes explicitly computing $\Psi_{\.c} = \frac{1}{m}\sum_{i=1}^m c_i y_i\Phi(\.x_i)$ and directly evaluating $\ip{\Psi_{\.c}}{\Psi_{\.c'}}$ as is done in SVM-perf.  Instead, one must compute  $J_{\.c\.c'} = \ip{\Psi_{\.c}}{\Psi_{\.c'}} = \frac{1}{m^2}\sum_{i,j=1}^m c_i c_j y_i y_j \ip{\Phi(\.x_i)}{\Phi(\.x_j)}$ as a sum of $O(m^2)$ inner products $\ip{\Phi(\.x_i)}{\Phi(\.x_j)}$, which are then each evaluated using the kernel trick. This rules out the possibility of an $O(m)$ algorithm, at least using methods that rely on the kernel trick to evaluate each $\ip{\Phi(\.x_i)}{\Phi(\.x_j)}$.  Noting that $\.w= \sum_{\.c}\alpha_{\.c}\Psi_{\.c}$ , the inner products $\ip{\.w}{\Phi(\.x_i)}$ are similarly expensive to compute directly classically if the dimension of the feature map is large.

\subsection{Quantum feature maps}

We now show how quantum computing can be used to efficiently approximate the inner products $\ip{\Psi_{\.c}}{\Psi_{\.c'}}$ and $\ip{\sum_{\.c}\alpha_{\.c}\Psi_{\.c}}{\Phi(\.x_i)}$, where high dimensional $\Psi_{\.c}$ can be implemented by a quantum circuit using only a number of qubits logarithmic in the dimension.  We first assume that the data vectors $\.x_i\in\mb{R}^d$ are encoded in the state of an $O(d)$-qubit register $\ket{\.x_i} $ via some suitable encoding scheme, e.g. given an integer $k$, $\.x_i$ could be encoded in $n = kd$ qubits by approximating each of the $d$ values of $\.x_i$ by a length $k$ bit string which is then encoded in a computational basis state of $n$ qubits. Once encoded, a quantum feature map encodes this information in larger space in the following way:

\begin{definitionenv}[Quantum feature map] Let $\+H_A = \lp \mb{C}^2\rp^{\otimes n}, \+H_B=\lp\mb{C}^2\rp^{\otimes N}$ be $n$-qubit input and $N$-qubit output registers respectively .  A quantum feature map is a unitary mapping $U_{\Phi}:\+H_A\otimes\+H_B\ra \+H_A\otimes\+H_B$ satisfying
\eq{
U_{\Phi}\ket{\.x}\ket{0}= \ket{\.x}\ket{\Phi(\.x)},
}
for each of the basis states $\.x\in\{0,1\}^n$, where $\ket{\Phi(\.x)} = \frac{1}{\norm{\Phi(\.x)}}\sum_{j=1}^{2^N} \Phi(\.x)_j \ket{j}$, with real amplitudes $\Phi(\.x)_j\in\mb{R}$. Denote the running time of $U_{\Phi}$ by $T_{\Phi}$. 
\end{definitionenv}
Note that the states $\ket{\Phi(\.x)}$ are not necessarily orthogonal. Implementing such a quantum feature map could be done, for instance, through a controlled parameterized quantum circuit.

We also define the quantum state analogy of $\Psi_{\.c}$ from Definition \ref{def:psi-c}:
\begin{definitionenv} Given a quantum feature map $U_{\Phi}$, define $\ket{\Psi_{\.c}}$ as
\eq{
\ket{\Psi_{\.c}} &= \frac{1}{\norm{\Psi_{\.c}}}\sum_{i=1}^m \frac{c_i y_i}{m} \norm{\Phi(\.x_i)}\ket{\Phi(\.x_i)}
}
where
\eq{
\norm{\Psi_{\.c}}^2 =\frac{1}{m^2}\sum_{i,j=1}^m c_ic_j y_i y_j \norm{\Phi(\.x_i)}\norm{\Phi(\.x_j)}\braket{\Phi(\.x_i)}{\Phi(\.x_j)}
}
\end{definitionenv}

\subsection{Quantum inner product estimation}

Let real vectors $x,y\in\mb{R}^{d}$ have corresponding normalized quantum states $\ket{x} = \frac{1}{\norm{x}}\sum_{i=1}^{d} x_i \ket{i}$ and $\ket{y} = \frac{1}{\norm{y}}\sum_{i=1}^{d} y_i \ket{i}$.  The following result shows how the inner product $\ip{x}{y} = \braket{x}{y}\norm{x}\norm{y}$ can be estimated efficiently on a quantum computer.

\begin{restatable}[Robust Inner Product Estimation \cite{allcock2018quantum}, restated]{theorem}{theoremripe}
\label{thm:ripe} 
Let $\ket{x}$ and $\ket{y}$ be quantum states with real amplitudes and with bounded norms $\norm{x},\norm{y}\le R$. If $\ket{x}$ and $\ket{y}$ can each be generated by a quantum circuit in time $T$, and if estimates of the norms are known to within $\epsilon/3R$ additive error, then one can perform the mapping $\ket{x}\ket{y}\ket{0}\rightarrow\ket{x}\ket{y}\ket{s}$ where, with probability at least $1-\delta$, $\abs{s-\< x,y\>} \leq  \epsilon $. The time required to perform this mapping is $\widetilde{O}\left(\frac{R^2\log(1/\delta)}{ \epsilon}T\right)$.
\end{restatable}

Thus, if one can efficiently create quantum states $\ket{\Psi_{\.c}}$ and estimate the norms $\norm{\Psi_{\.c}}$, then the corresponding $J_{\.c\.c'} = \ip{\Psi_{\.c}}{\Psi_{\.c'}} = \norm{\Psi_{\.c}}\norm{\Psi_{\.c'}}\braket{\Psi_{\.c}}{\Psi_{\.c'}}$ can be approximated efficiently. In this section we show that this is possible with a quantum random access memory (qRAM), which is a device that allows classical data to be queried efficiently in superposition. That is, if $x\in\mb{R}^d$ is stored in qRAM, then a query to the qRAM implements the unitary $\sum_j \alpha_j\ket{j}\ket{0}\ra \sum_j \alpha_j\ket{j}\ket{x_j}$.  If the elements $x_j$ of $x$ arrive as a stream of entries $(j, x_j)$ in some arbitrary order, then $x$ can be stored in a particular data structure \cite{kerenidis2017quantum} in time $\tilde{O}(d)$ and, once stored, $\ket{x} = \frac{1}{\norm{x}}\sum_j x_j\ket{j}$ can be created in time polylogarithmic in $d$.  Note that when we refer to real-valued data being stored in qRAM, it is implied that the information is stored as a binary representation of the data, so that it may be loaded into a qubit register.

\begin{restatable}[]{theorem}{jip}
\label{thm:q-ip-epsilon}
 Let $\.c,\.c' \in \{0,1\}^m$. If, for all $i\in[m]$, $\.x_i,\frac{c_i y_i}{\sqrt{m}}\norm{\Phi(\.x_i)}, \frac{c'_i y_i}{\sqrt{m}}\norm{\Phi(\.x_i)}$ are stored in qRAM, and if $\eta_{\.c}= \sqrt{\frac{\sum_{i=1}^m c_i\norm{\Phi(\.x_i)}^2}{m}}$ and $\eta_{\.c'}= \sqrt{\frac{\sum_{i=1}^m c'_i\norm{\Phi(\.x_i)}^2}{m}}$ are known then, with probability at least $1-\delta$, an estimate $s_{\.c\.c'}$ satisfying
\eq{
\abs{s_{\.c\.c'} - \ip{\Psi_{\.c}}{\Psi_{\.c'}}} &\le \epsilon 
}
can be computed in time 
\eql{T_{cc'} &= \tilde{O}\lp 
\frac{\log(1/\delta)}{\epsilon }\frac{R^3}{\min\left\{\norm{\Psi_{\.c}},\norm{\Psi_{\.c'}}\right\}}T_\Phi\rp
\label{eq:time-J-estimate}
}
where  $R=\max_i \norm{\Phi(\.x_i)}$.
\end{restatable}

A similar result applies to estimating inner products of the form $\sum_{\.c \alpha_c}\ip{ \Psi_{\.c}}{y_i\Phi(\.x_i)}$.

\begin{restatable}[]{theorem}{zetaip}
\label{thm:q-ip-epsilon-2}
Let $\+W \subseteq \{0,1\}^m$ and $\sum_{\.c\in\+W}\alpha_{\.c}\le C$. If $\eta_{\.c}$ are known for all $\.c\in\+W$ and if  $\.x_i$ and $\frac{c_i y_i}{\sqrt{m}}\norm{\Phi(\.x_i)}$ are stored in qRAM for all $i\in[m]$ then, with probability at least $ 1 - \delta$, $\sum_{\.c\in\+W}\alpha_{\.c} \ip{\Psi_{\.c}}{y_i\Phi(\.x_i)}$ can be estimated to within error $\epsilon$ in time 
\eq{
\tilde{O}\lp \frac{\log(1/\delta)}{\epsilon }
\frac{CR^3\abs{\+W} }{\min_{\.c\in\+W} \norm{\Psi_{\.c}} } T_\Phi\rp
}
\end{restatable}

Proofs of Theorems \ref{thm:q-ip-epsilon} and \ref{thm:q-ip-epsilon-2} are given in Appendix \ref{app:method-proofs}.


\section{Linear Time Algorithm for Nonlinear SVMs}

The results of the previous section can be used to generalize Joachims' algorithm to quantum feature-mapped data. Let $S^+_n$ denote the cone of $n\times n$ positive semi-definite matrices. Given $X\in\mb{R}^{n\times n}$, let $P_{S^+_n}(X) = \arg\min_{Y\in S^+_n} \norm{Y - X}_F$, i.e. the projection of $X$ onto $S^+_n$, where $\norm{\cdot}_F$ is the Frobenius norm. Denote the $i$-th row of $X$ by $\lp X\rp_i$.  

Define $IP_{\epsilon, \delta}(x,y)$ to be a quantum subroutine which, with probability at least $1-\delta$, returns an estimate $s$ of the inner product of two vectors $x,y$ satisfying $\abs{s - \ip{x}{y}}\le \epsilon$.  As we have seen, with appropriate data stored in qRAM, this subroutine can be implemented efficiently on a quantum computer.

Our quantum algorithm for nonlinear structural SVMs is presented in Algorithm \ref{alg:qc-struct-svm}. At first sight, it appears significantly more complicated than Algorithm \ref{alg:struct-svm}, but this is due in part to more detailed notation used to aid the analysis later.  The key differences are (i) the matrix elements $J_{\.c\.c'} = \ip{\Psi_{\.c}}{\Psi_{\.c'}}$ are only estimated to precision $\epsilon_{J}$ by the quantum subroutine; (ii)
as the corresponding matrix $J$ is not guaranteed to be positive semi-definite, an additional classical projection step must therefore be carried out to map the estimated matrix on to the p.s.d. cone at each iteration; (iii) In the classical algorithm, the values of $c^*_i$ are deduced by $c_i^* = \max(0, 1 - y_i(\.w^T \.x_i))$ whereas here we can only estimate the inner products $\ip{\.w^T}{\Phi(\.x_i)}$ to precision $\epsilon$, and $\.w$ is known only implicitly according to $\.w = \sum_{\.c\in\+W}\alpha_{\.c}\Psi_{\.c}$. Note that apart from the quantum inner product estimation subroutines, all other computations are performed classically. 

\begin{Algorithm*}
\captionsetup{labelsep=newline,format=hline_caption}

\caption{Quantum-classical structural SVM algorithm}

\label{alg:qc-struct-svm}
\begin{algorithmic}
\STATE  \textbf{Input: }Training set $S = \left\{(\.x_1, y_1), \ldots, (\.x_m, y_m)\right\}$, SVM hyperparameter $C$,  quantum feature map $U_{\Phi}$ with maximum norm $R = \max_i\norm{{\Phi(\.x_i)}}$, tolerance parameters $\epsilon, \delta > 0$, $\.c\in\{0,1\}^m$, $t_{\max} \ge 1$.
\STATE
\STATE set $t \gets 1$ and $\+W_1 \gets \left\{ \.c \right\}$
\STATE
\FOR{$i=1,\ldots, m$}
\STATE Store $\frac{c_iy_i}{\sqrt{m}}\norm{\Phi(\.x_i)}$ and $\.x_i$ in qRAM
\ENDFOR
\STATE Compute and store $\eta_{\.c} = \sqrt{\frac{\sum_{i=1}^m c_i\norm{\Phi(\.x_i)}^2}{m}}$ classically
\STATE
\REPEAT
\STATE set $\epsilon_{J} \gets \frac{1}{Ct\,t_{\max}}$ and $\delta_J \gets \frac{\delta}{2t^2 \,t_{\max}}$
\STATE
\FOR{$\.c,\.c'\in \+W_t$} 
    \STATE $\tilde{J}_{\.c\.c'}\gets IP_{\epsilon_{J},\delta_J}(\Psi_{\.c},\Psi_{\.c'})$ 
\ENDFOR
    \STATE
    \STATE $\hat{J}_{\+W_t}\gets P_{S^+_{\abs{\+W_t}}}(\tilde{J})$
    \STATE $\hat{\alpha}^{(t)}\gets \argmax_{\alpha \ge 0} -\frac{1}{2}\sum_{\.c,\.c'\in{\+W}}\alpha_{\.c}\alpha_{\.c'} \lp \hat{J}_{\+W_t}\rp_{\.c\.c'} + \sum_{\.c\in\+W} \frac{\norm{\.c}_1}{m}\alpha_{\.c}$
    \STATE $\quad\quad\quad\quad\quad\quad\quad \text{s.t.}\quad \sum_{\.c\in\+W_t}\alpha_{\.c}\le C$
    \STATE Store $\hat{\alpha}^{(t)}$ in qRAM
\STATE
\FOR{$\.c\in\+W_t$}
    \STATE $\xi_{\.c}^{(t)}\gets \max\left\{\frac{1}{m}\sum_{i=1}^m c_i - \sum_{\.c'\in\+W_t} \hat{\alpha}^{(t)}_{\.c'}\lp\hat{J}_{\+W_t}\rp_{\.c\.c'}, 0 \right\}$
\ENDFOR
\STATE
\STATE set $\hat{\xi}^{(t)} \gets \max_{\.c\in\+W_t} \xi_{\.c}^{(t)} + \frac{1}{t_{\max}}$ and $\delta_{\zeta} \gets \frac{\delta}{2m\,t_{\max}}$
\STATE
\FOR{$i=1,\ldots, m$}
    \STATE $\zeta_i \gets IP_{\epsilon,\delta_\zeta} \lp \sum_{\.c\in\+W_t} \hat{\alpha}^{(t)}_{\.c}\Psi_{\.c},y_i\Phi(\.x_i)\rp$
    \STATE $c^{(t+1)}_i \gets \begin{cases}1 & \zeta_i  < 1\\ 0 & \text{otherwise}\end{cases}$
    \STATE Store $\frac{c_i^{(t+1)}y_i}{\sqrt{m}}\norm{\Phi(\.x_i)}$ in qRAM
\ENDFOR
\STATE
\STATE Compute and store $\eta_{\.c^{(t+1)}}$ classically
\STATE set $\+W_{t+1}\gets \+W_t\cup \{\.c^{(t+1)}\}$ and $t\gets t+1$
\STATE
\UNTIL{$\frac{1}{m}\sum_{i=1}^m\max \left\{0, 1 - \zeta_i\right\} \le \hat{\xi}^{(t)} + 2\epsilon $ OR $t > t_{\max}$}
\STATE
\STATE \textbf{Output: }$\hat{\alpha} = \hat{\alpha}^{(t)}, \hat{\xi} = \hat{\xi}^{(t)}$ 
\end{algorithmic}
\hrulefill
\end{Algorithm*}

\begin{restatable}[]{theorem}{algits}\label{thm:hybrid-algo-iterations}
Let $t_{\max}$ be a user-defined parameter and let $(\.w^*, \xi^*)$ be an optimal solution of OP \ref{op:struct-svm-p}. If Algorithm \ref{alg:qc-struct-svm} terminates in at most $t_{\max}$ iterations then, with probability at least $1-\delta$, it outputs $\hat{\alpha}$ and $\hat{\xi}$ such that $\hat{\.w}= \sum_{\.c} \hat{\alpha}_{\.c} \Psi_{\.c}$ satisfies  $P(\hat{\.w}, \hat{\xi}) - P(\.w^*,\xi^*) \le \min\left\{\frac{C\epsilon}{2},\frac{\epsilon^2}{8R^2}\right\}$, and  $(\hat{\.w}, \hat{\xi} + 3\epsilon)$ is feasible for OP \ref{op:struct-svm-p}.  If $t_{\max}\ge\max\left\{\frac{4}{\epsilon}, \frac{16CR^2}{\epsilon^2}\right\}$ then the algorithm is guaranteed to terminate in at most $t_{\max}$ iterations. 
\end{restatable}

\begin{proof}
See Appendix \ref{app:proof-of-algo}.
\end{proof}

\begin{restatable}[]{theorem}{algtime}\label{thm:hybrid-algo-time}
Algorithm $\ref{alg:qc-struct-svm}$ has a time complexity of 
 \eql{
\tilde{O}\lp \frac{CR^3\log(1/\delta)}{\Psi_{\min}} \lp   \frac{t^2_{\max}}{\epsilon}\cdot m+ t_{\max}^5 \rp T_\Phi  \rp \label{eq:time-scaling}
}
where $\Psi_{\min} = \min_{\.c\in\+W_{t_f}}\norm{\Psi_{\.c}}$, and $t_f\le t_{\max}$ is the iteration at which the algorithm terminates.
\end{restatable}

\begin{proof}
See Appendix \ref{app:hybrid-algo-time}.

\end{proof}

The total number of outer-loop iterations (indexed by $t$) of Algorithm $\ref{alg:qc-struct-svm}$ is upper-bounded by the choice of $t_{\max}$. One may wonder why we do not simply set $t_{\max} = \max\left\{\frac{4}{\epsilon}, \frac{16CR^2}{\epsilon^2}\right\}$ as this would ensure that, with high probability, the algorithm outputs a nearly optimal solution. The reason is that $t_{\max}$ also affects the the quantities $\epsilon_{J} = \frac{1}{Ctt_{\max}}$, $\delta_J = \frac{\delta}{2t^2 t_{\max}}$ and $\delta_\zeta = \frac{\delta}{2mt_{\max}}$.  These in turn impact the running time of the two quantum inner product estimation subroutines that take place in each iteration, e.g. the first quantum inner product estimation subroutine has running time that scales like $\frac{\log(1/\delta_J)}{\epsilon_J}$. While the upper-bound on $t_{\max}$ of $\max\left\{\frac{4}{\epsilon}, \frac{16CR^2}{\epsilon^2}\right\}$ is independent of $m$, it can be large for reasonable values of the other algorithm parameters $C$, $\epsilon$, $\delta$ and $R$. For instance, the choice of $(C,\epsilon,\delta,R) = (10^4, 0.01, 0.1, 1)$ which, as we show in the Simulation section, lead to good classification performance on the datasets we consider, corresponds to $t_{\max} = 1.6\times 10^{9}$, and $\frac{\log(1/\delta_J)}{\epsilon_J} \ge 1.6\times 10^{13}$.  In practice, we find that this upper-bound on $t_{\max}$ is very loose, and the situation is far better in practice: the algorithm can terminate successfully in very few iterations with much smaller values of $t_{\max}$. In the examples we consider, the algorithm terminates successfully before $t$ reaches $t_{\max}=50$, corresponding to $\frac{\log(1/\delta_J)}{\epsilon_J} \le 3.7\times 10^8$.

The running time of Algorithm $\ref{alg:qc-struct-svm}$ also depends on the quantity $\Psi_{\min}$ which is a function of both the dataset as well as the quantum feature map chosen.  While this can make $\Psi_{\min}$ hard to predict, we will again see in the Simulation section that in practice the situation is optimistic: we empirically find that $\Psi_{\min}$ is neither too small, nor does it scale noticeably with $m$ or the dimension of the quantum feature map.

\subsection{Classification of new test points}

As is standard in SVM theory, the solution $\hat{\alpha}$ from Algorithm \ref{alg:qc-struct-svm} can be used to classify a new data point $\.x$ according to
\eq{
y_{pred} = \sgn \ip{\sum_{\.c}\hat{\alpha}_{\.c}\Psi_{\.c}}{ \Phi(\.x)}
}
where $y_{pred}$ is the predicted label of $\.x$. From Theorem \ref{thm:q-ip-epsilon-2}, and noting that $\abs{\+W}\le t_{\max}$, we obtain the following result:

\begin{restatable}[]{theorem}{qclas}\label{thm:q-classification} Let $\hat{\alpha}$ be the output of Algorithm \ref{alg:qc-struct-svm}, and let $\.x$ be stored in qRAM. There is a quantum algorithm that, with probability at least $1-\delta$, estimates the inner product $\ip{\sum_{\.c}\hat{\alpha}_{\.c}\Psi_{\.c}}{ \Phi(\.x)}$ to within error $\epsilon$ in time $\tilde{O}\lp\frac{C R^3 \log(1/\delta)}{\Psi_{\min}}\frac{t_{\max}}{\epsilon}T_{\Phi}\rp$
\end{restatable} 
Taking the sign of the output then completes the classification.


\section{Simulation}

While the true performance of our algorithms for large $m$ and high dimensional quantum feature maps necessitate a fault-tolerant quantum computer to evaluate, we can gain some insight into how it behaves by performing smaller scale numerical experiments on a classical computer.  In this section we empirically find that the algorithm can have good performance in practice, both in terms of classification accuracy as well as in terms of the parameters which impact running time. 

\subsection{Data set}

\begin{figure}
\centering
\includegraphics[height=5.0cm]{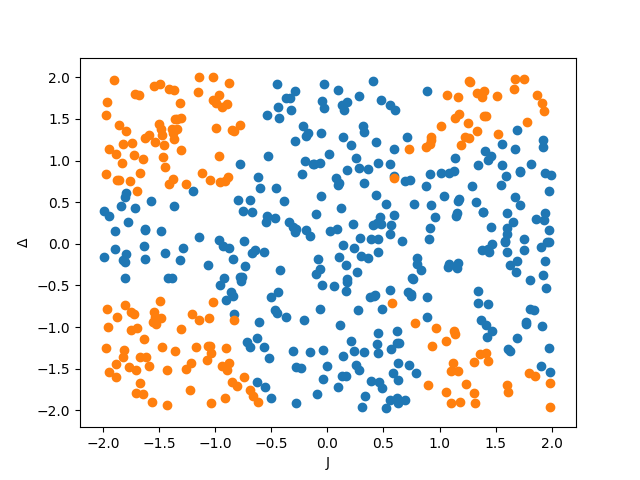}
\includegraphics[height=5.0cm]{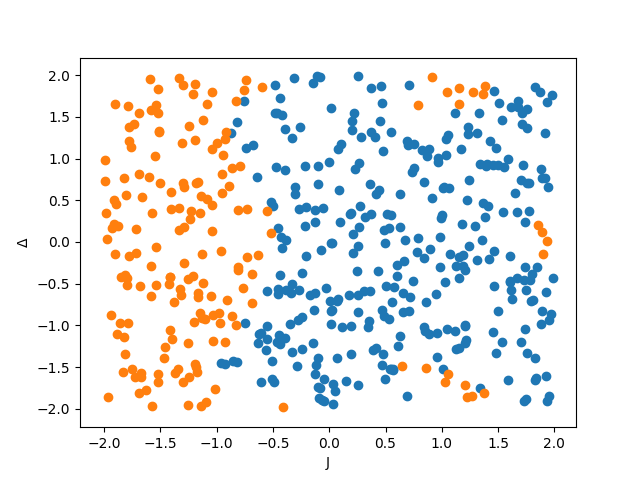}
\caption{Sample data sets (left) $S_{5,500}(1.5,J,1,\Delta, 3,0.5)$ and (right) $S_{8,500}(2.4,J,1,\Delta, 2,0.5)$. Blue and orange colors indicated $+1$ and $-1$ labels respectively.  In each case $500$ data points $(J,\Delta)$ were generated uniformly at random in the range $[-2,2]^2$.}
\label{fig:data-sample}
\end{figure}

To test our algorithm we need to choose both a data set as well as a quantum feature map. The general question of what constitutes a good quantum feature map, especially for classifying classical data sets, is an open problem and beyond the scope of this investigation.  However, if the data is generated from a quantum problem, then physical intuition may guide our choice of feature map. We therefore consider the following toy example which is nonetheless instructive. Let $H_N$ be the Hamiltonian of a generalized Ising Hamiltonian on $N$ spins 
\eql{
H_N(\vec{J},\vec{\Delta},\vec{\Gamma}) &= - \sum_{j=1}^N J_j Z_j\otimes Z_{j+1} + \sum_{j=1}^N \lp \Delta_j X_j + \Gamma_j Z_j\rp \label{eq:ising-general}
}
where $\vec{J},\vec{\Delta},\vec{\Gamma}$ are vectors of real parameters to be chosen, and $Z_j, X_j$ are Pauli $Z$ and $X$ operators acting on the $j$-th qubit in the chain, respectively. 
We generate a data set by randomly selecting $m$ points $(\vec{J},\vec{\Delta},\vec{\Gamma})$ and labelling them according to whether the expectation value of the operator $M = \frac{1}{N}\lp \sum_j Z_j\rp^2$ with respect to the ground state of $H_N(\vec{J},\vec{\Delta},\vec{\Gamma})$ satisfies
\eql{
\langle M\rangle &\quad
\begin{cases}
\ge \mu_0 &  (+1 \text{ labels}) \\
< \mu_0 & (-1 \text{ labels}) 
\end{cases}\label{eq:mag-mu}
}
for some cut-off value $\mu_0$, i.e. the points are labelled depending on whether the average total magnetism squared is above or below $\mu_0$.  In our simulations we consider a special case of \eqref{eq:ising-general} where $J_j = J\cos \frac{k_J \pi (j-1)}{N}$, $\Delta_j = \Delta \sin\frac{k_\Delta \pi j}{N}$ and $\Gamma_j = \Gamma$, where $J, k_J, \Delta, k_\Delta, \Gamma$ are real.  Examples of data sets $S_{N,m}$ corresponding to such a Hamiltonian, whose parameters we notate by 
\eq{
S_{N,m}(\mu_0,J,k_J,\Delta, k_{\Delta},\Gamma),
}
can be found in Fig \ref{fig:data-sample}.

\subsection{Quantum feature map}
For quantum feature map we choose
\eql{
\ket{\Psi(\vec{J},\vec{\Delta},\vec{\Gamma})} &= \frac{\ket{0}\ket{0}\ket{0} + \ket{1}\ket{\psi_{GS}}\ket{\psi_{GS}}}{\sqrt{2}} \label{eq:q-feature-map}
}
where $\ket{\psi_{GS}}$ is the ground state of \eqref{eq:ising-general} and, as it is a normalized state, has corresponding value of $R=1$. We compute such feature maps classically by explicitly diagonalizing $H_N$. In a real implementation of our algorithm on a quantum computer, such a feature map would be implemented  by a controlled unitary for generating the (approximate) ground state of $H_N$, which could be done by a variety of methods e.g. by digitized adiabatic evolution or methods based on  imaginary time evolution \cite{motta2019quantum, hsieh2019rbm}, with running time $T_{\phi}$ dependent on the degree of accuracy required.   The choice of \eqref{eq:q-feature-map} is motivated by noting that condition \eqref{eq:mag-mu} is equivalent to determining the sign of $\ip{W}{\Psi}$, where $W$ is a vector which depends only on $\mu_0$, and not on the choice of parameters in $H_N$ (see Appendix \ref{app:feature-map}). By construction, $W$ defines a separating hyperplane for the data, so the chosen quantum feature map separate the data in feature space. As the Hamiltonian is real, it has a set of real eigenvectors and hence $\ket{\Psi}$ can be defined to have real amplitudes, as required.
 
 \subsection{Numerical results}
 
We first evaluate the performance of our algorithm on data sets $S_{N,m}$ for $N=6$ and increasing orders of $m$ from $10^2$ to $10^5$.  
\begin{itemize}
    \item For each value of $m$, a data set $S_{6,m}(\mu_0,J,k_J,\Delta,k_{\Delta},\Gamma)$ was generated for points $(J,\Delta)$ sampled uniformly at random in the range $[-2,2]^2$.
    \item The values of $\mu_0, k_j, k_\Delta,\Gamma$ were fixed and chosen to give roughly balanced data, i.e. the ratio of $+1$ to $-1$ labels is no more than 70:30 in favor of either label.
   \item Each set of $m$ data points was divided into training and test sets in the ratio 70:30, and training was performed according to Algorithm $\ref{alg:qc-struct-svm}$ with parameters $(C,\epsilon,\delta,t_{\max})= (10^4, 10^{-2}, 10^{-1},50)$.  
    \item These values of $C$ and $\epsilon$ were selected to give classification accuracy competitive with classical SVM algorithms utilizing standard Gaussian radial basis function (RBF) kernels, with hyperparameters trained using a subset of the training set of size $20\%$ used for hold-out validation. Note that the quantum feature maps do not have any similar tunable parameters, and a modification of \eqref{eq:q-feature-map}, for instance to include a tunable weighting between the two parts of the superposition, could be introduced to further improve performance.
    \item The quantum $IP_{\epsilon,\delta}$ inner product estimations in the algorithm were approximated by adding Gaussian random noise to the true inner product, such that the resulting inner product was within $\epsilon$ of the true value with probability at least $1-\delta$.  Classically simulating quantum $IP_{\epsilon,\delta}$ inner product estimation with inner products distributed according to the actual quantum procedures underlying Theorems \ref{thm:q-ip-epsilon} and \ref{thm:q-ip-epsilon-2} was too computationally intensive in general to perform. However, these were tested on small data sets and quantum feature vectors, and found to behave very similarly to adding Gaussian random noise. This is consistent with the results of the numerical simulations in \cite{allcock2018quantum}.
\end{itemize}
  Note that the values of $C,\epsilon,\delta$ chosen correspond to $\max\left\{\frac{4}{\epsilon}, \frac{16CR^2}{\epsilon^2}\right\} > 10^9$. This is an upper-bound on the number of iterations $t_{\max}$ needed for the algorithm to converge to a good solution.  However, we find empirically that $t_{\max} = 50$ is sufficient for the algorithm to terminate with a good solution across the range of $m$ we consider. 
 
 The results are shown in Table \ref{tb:example-data-01}.  We find that (i) with these choices of $C,\epsilon,\delta,t_{\max}$  our algorithm has high classification accuracy, competitive with standard classical SVM algorithms utilizing RBF kernels with optimized hyperparameters. (ii)  $\Psi_{\min}$ is of the order $10^{-2}$ in these cases, and does scale noticeably over the range of $m$ from $10^2$ to $10^5$. If $\Psi_{\min}$ were to decrease polynomially (or worse, exponentially) in $m$ then this would be a severe limitation of our algorithm.  Fortunately this does not appear to be the case.
 
 \begin{table}
\begin{tabular}{c|cccc}
     $m$ & $10^2$  & $10^3$  &  $10^4$  & $10^5$ \\ \hline
     $\Psi_{\min}$ & 0.010 & 0.018 & 0.016 & 0.011\\ 
     iterations & 36 & 38 &  39 &  38 \\
     accuracy $(\%)$ & 93.3 & 99.3  & 99.0  & 98.9  \\ \hline
     RBF accuracy $(\%)$ & 86.7 & 96.0 & 99.0 & 99.7 
\end{tabular}
\caption{$\Psi_{\min}$, iterations $t$ until termination, and classification accuracy of Algorithm $\ref{alg:qc-struct-svm}$ on data sets with parameters $S_{6,m}(1.8,J,1,\Delta, 9,0.2)$ for $m$ randomly chosen points $(J,\Delta)\in[-2,2]^2$, with $m$ in the range $m=10^2$ to $m=10^5$.  Algorithm parameters were chosen to be $(C,\epsilon,\delta,t_{\max})= (10^4, 10^{-2}, 10^{-1},50)$. The classification accuracy of classical SVMs with Gaussian radial basis function kernels and optimized hyperparameters is given for comparison.}\label{tb:example-data-01}
\end{table}

 We further investigate the behaviour of $\Psi_{\min}$ by generating data sets $S_{N,m}$ for fixed $m=1000$ and $N$ ranging from $4$ to $8$.  For each $N$, we generate $100$ random data sets $S_{N,m}(\mu_0, J, k_J, \Delta,k_\Delta, \Gamma)$, where each data set consists of $1000$ points $(J,\Delta)$ sampled uniformly at random in the range $[-2,2]^2$, and random values of $\mu_o, k_J, k_\Delta, \Gamma$ chosen to give roughly balanced data sets as before.  Unlike before, we do not divide the data into training and test sets. Instead, we perform training on the entire data set, and record the value of $\Psi_{\min}$ in each instance.  The results are given in Table \ref{tb:psi-min-N} and show that across this range of $N$ (i) the average value  $\bar{\Psi}_{\min}$ is of order $10^{-2}$ (ii) the spread around this average is fairly tight, and the minimum value of $\Psi_{\min}$ in any single instance is of order $10^{-3}$.  These support the results of the first experiment, and indicate that the value of $\Psi_{\min}$ may not adversely affect the running time of the algorithm in practice.

\begin{table}
\begin{tabular}{c|ccccc}
     $N$  & $4$  & 5 & $6$  & 7 & $8$ \\ \hline
     $\bar{\Psi}_{\min}(10^{-2})$  & 1.28 & 1.34 & 1.32 & 1.44 &  1.16\\
    $\min(\Psi_{\min})$ $(10^{-3})$ & 3.41 & 2.33 & 3.16 & 3.82 & 3.96 \\
     s.d. ($10^{-3}$)  & 8.6 & 20.4  & 16.2 &  19.6 & 6.4 \\
\end{tabular}
\caption{Average value, minimum value, and standard deviation of $\Psi_{\min}$ for random data sets generated $S_{N,1000}(\mu_0,J,k_J,\Delta, k_{\Delta},\Gamma)$.  For each value of $N$, 100 instances (of $m=1000$ data points each) were generated for random values of $k_J, k_\Delta$ and $\Gamma$, with $\mu_0 = 3N$ .  Algorithm  $\ref{alg:qc-struct-svm}$ was trained using parameters $(C,\epsilon,\delta,t_{\max})= (10^4, 10^{-2}, 10^{-1},50)$.}\label{tb:psi-min-N}
\end{table}


\section{Conclusions}

We have proposed a quantum extension of SVM-perf for training nonlinear soft-margin $\ell_1$-SVMs in time linear in the number of training examples $m$, up to polylogarithmic factors, and given numerical evidence that the algorithm can perform well in practice as well as in theory.  This goes beyond classical SVM-perf, which achieves linear $m$ scaling only for linear SVMs or for feature maps corresponding to low-rank or shift-invariant kernels, and brings the theoretical running time and applicability of SVM-perf in line with the classical Pegasos algorithm which --- in spite of having best-in-class asymptotic guarantees --- has empirically been outperformed by other methods on certain datasets. Our algorithm also goes beyond previous quantum algorithms which achieve linear or better scaling in $m$ for other variants of SVMs, which lack some of the desirable properties of the soft-margin $\ell_1$-SVM model. Following this work, it is straightforward to propose a quantum extension of Pegasos. An interesting question to consider is how such an algorithm would perform against the quantum SVM-perf algorithm we have presented here.

Another important direction for future research is to investigate methods for selecting good quantum feature maps and associated values of $R$ for a given problem. While work has been done on learning quantum feature maps by training parameterizable quantum circuits \cite{romero2017quantum, farhi2018classification, havlivcek2019supervised, Lloyd2020embeddings}, a deeper understanding of quantum feature map construction and optimization is needed. In particular, the question of when an explicit quantum feature map can be advantageous compared to the classical kernel trick -- as implemented in Pegasos or other state-of-the-art algorithms -- needs further investigation.  Furthermore, in classical SVM training, typically one of a number of flexible, general purpose kernels such as the Gaussian RBF kernel can be employed in a wide variety of settings. Whether similar, general purpose quantum feature maps can be useful in practice is an open problem, and one that could potentially greatly affect the adoption of quantum algorithms as a useful tool for machine learning.

\section{Acknowledgements}
We are grateful to Shengyu Zhang for many helpful discussions and feedback on the manuscript.


\newpage
\bibliographystyle{unsrtnat}
\bibliography{quantum-kernel.bib}

\onecolumn\newpage
\appendix

\section{Proofs of Theorem \ref{thm:q-ip-epsilon} and Theorem \ref{thm:q-ip-epsilon-2}} \label{app:method-proofs}

\begin{lem}\label{lem:create-psi} If $\.x_1,\ldots , \.x_m \in\mb{R}^d$ and $\frac{c_1 y_1}{\sqrt{m}}\norm{\Phi(\.x_1)}, \ldots ,\frac{c_m y_m}{\sqrt{m}}\norm{\Phi(\.x_m)}$ for $\.c\in\{0,1\}^m$ are stored in qRAM, and if $\eta_{\.c}= \sqrt{\frac{\sum_{i=1}^m c_i\norm{\Phi(\.x_i)}^2}{m}}$ is known, then  $\ket{\Psi_{\.c}}$ can be created in time 
$T_{\Psi_{\.c}} = \tilde{O}\lp \frac{R}{\norm{\Psi_{\.c}}}T_{\Phi} \rp$, and $\norm{\Psi_{\.c}}$ estimated to additive error $\epsilon/3R$ in time $O\lp \frac{R^3}{\epsilon}T_{\Phi}\rp$


\end{lem}

\begin{proof}


With the above values in qRAM, unitary operators $U_{\.x}$ and $U_{\.c}$ can be implemented in times $T_{U_{\.x}} = \polylog(md)$ and $T_{U_{\.c}} = \polylog(m)$, which effect the transformations
\eq{
U_{\.x}\ket{i}\ket{0}&=\ket{i}\ket{\.x_i}\\
U_{\.c}\ket{0}&= \frac{1}{\eta_{\.c}}\sum_{j=0}^{m-1} \frac{c_j y_j}{\sqrt{m}} \norm{\Phi(\.x_j)}\ket{j}
}
$\ket{\Psi_c}$ can then be created by the following procedure:

\eql{
\ket{0}\ket{0}\ket{0}&\xra{U_{\.c}}\frac{1}{\eta_c}\sum_{j=0}^{m-1} \frac{c_j y_j}{\sqrt{m}} \norm{\Phi(\.x_j)}\ket{j}\ket{0}\ket{0} \nonumber\\
&\xra{U_{\.x}}\frac{1}{\eta_c}\sum_{j=0}^{m-1} \frac{c_j y_j}{\sqrt{m}}\norm{\Phi(\.x_j)}\ket{j}\ket{\.x_j}\ket{0} \nonumber\\
&\xra{U_{\Phi}}\frac{1}{\eta_c}\sum_{j=0}^{m-1} \frac{c_j y_j}{\sqrt{m}}\norm{\Phi(\.x_j)}\ket{j}\ket{\.x_j}\ket{\Phi(\.x_j)} \nonumber\\
&\xra{U_{\.x}^\dag}\frac{1}{\eta_c}\sum_{j=0}^{m-1} \frac{c_j y_j}{\sqrt{m}}\norm{\Phi(\.x_j)}\ket{j}\ket{0}\ket{\Phi(\.x_j)} \nonumber
}
Discarding the $\ket{0}$ register, and applying the Hadamard transformation $H\ket{j} = \frac{1}{\sqrt{m}}\sum_k (-1)^{j\cdot k }\ket{k}$ to the first register then gives

\eql{
&\xra{\text{H}} \frac{1}{\eta_{\.c}}\sum_{j=0}^{m-1} \frac{c_j y_j}{\sqrt{m}}\norm{\Phi(\.x_j)}\frac{1}{\sqrt{m}}\sum_{k=0}^{m-1}(-1)^{j\cdot k}\ket{k}\ket{\Phi(\.x_j)} \nonumber\\
&= \frac{\norm{\Psi_{\.c}}}{\eta_{\.c}}\ket{0}\frac{1}{\norm{\Psi_{\.c}}}\sum_{j=0}^{m-1} \frac{c_j y_j}{m}\norm{\Phi(\.x_j)}\ket{\Phi(\.x_j)} + \ket{0^\perp, \text{junk}} \nonumber\\
&= \frac{\norm{\Psi_{\.c}}}{\eta_{\.c}}\ket{0}\ket{\Psi_{\.c}} + \ket{0^\perp, \text{junk}} \label{eq:psi-plus-junk}
}
where $\ket{0^\perp, \text{junk}}$ is an unnormalized quantum state where the first qubit is orthogonal to $\ket{0}$. The state $\frac{\norm{\Psi_{\.c}}}{\eta_{\.c}}\ket{0}\ket{\Psi_{\.c}} + \ket{0^\perp,\text{junk}}$ can therefore be created in time $ T_{U_{\.c}}+ 2T_{U_{\.x}}+ T_\Phi = \tilde{O}(T_\Phi)$. 

By quantum amplitude amplification and amplitude estimation \cite{brassard2002quantum}, given access to a unitary operator $U$ acting on $k$ qubits such that $U\ket{0}^{\otimes k} = \sin(\theta) \ket{x, 0} + \cos(\theta) \ket{G, 0^\perp}$ (where $\ket{G}$ is arbitrary),  $\sin^{2}(\theta)$ can be estimated to additive error $\epsilon$  in time $O\lp \frac{T(U)}{\epsilon}\rp$ and 
$\ket{x}$ can be generated in expected time $O\lp \frac{T(U)}{\sin (\theta)}\rp$,where $T(U)$ is the time required to implement $U$. 
Amplitude amplification applied to the unitary creating the state in \eqref{eq:psi-plus-junk} allows one to create $\ket{\Psi_{\.c}}$ in expected time $\tilde{O}\lp \frac{\eta_{\.c}}{\norm{\Psi_{\.c}}}T_\Phi\rp = \tilde{O}\lp \frac{R}{\norm{\Psi_{\.c}}}T_\Phi\rp$, since $\eta_c \le \sqrt{\frac{\sum_{i=1}^m\norm{\Phi(\.x_i)^2}}{m}} \le R$.  Similarly, amplitude estimation can be used to obtain a value $s$ satisfying $\abs{s - \frac{\norm{\Psi_{\.c}^2}}{\eta_{\.c}^2}}
\le \frac{\epsilon}{3R^3}$ 
 in time $\tilde{O}\lp \frac{R^3}{\epsilon}T_\Phi\rp$.  Outputting $\overline{\norm{\Psi_{\.c}}} = \eta_{\.c}^2s$ then satisfies $\abs{\overline{\norm{\Psi_{\.c}}} - \norm{\Psi_{\.c}}} \le\frac{\epsilon}{3R}$.

\end{proof}

\jip*

\begin{proof}
From Lemma \ref{lem:create-psi}, the states $\ket{\Psi_{\.c}}$ and $\ket{\Psi_{\.c'}}$ can be created in time $\tilde{O}\lp \frac{R}{\min\left\{\norm{\Psi_{\.c}},\norm{\Psi_{\.c'}}\right\}}T_\Phi\rp$, and estimates of their norms to $\epsilon /3R$ additive error can be obtained in time $\tilde{O}\lp\frac{R^3}{\epsilon}T_\Phi\rp $. From Theorem \ref{thm:ripe} it follows that an estimate $s_{cc'}$ satisfying
\eq{
\abs{s_{\.c\.c'} - \ip{\Psi_{\.c}}{\Psi_{\.c'}}} &\le  \epsilon 
}
can be found with probability at least $1-\delta$ in time 
\eq{
T_{est} = \tilde{O}\lp 
\frac{\log(1/\delta)}{\epsilon }\frac{R^3}{\min\left\{\norm{\Psi_{\.c}},\norm{\Psi_{\.c'}}\right\}}T_\Phi\rp
}
\end{proof}

\zetaip*

\begin{proof}

With the above data in qRAM, an almost identical analysis to that in Theorem \ref{thm:q-ip-epsilon} can be applied to deduce that, for any $\.c\in\+W$, with probability at least $ 1-\delta/\abs{\+W}$,  an estimate $t_{\.c i}$ satisfying

\eq{
\abs{t_{\.c i} - \ip{\Psi_{\.c}}{y_i\Phi(\.x_i)}} &\le \epsilon/C
}
can be computed in time 
\eq{
T_{\.c i} &= \tilde{O}\lp 
\frac{C\log(\abs{\+W}/\delta)}{\epsilon }\frac{R^3}{\min_{\.c\in\+W} \norm{\Psi_{\.c}}}  T_\Phi\rp 
}
and the total time required to estimate all $\abs{\+W}$ terms (i.e. $t_{\.ci}$ for all $\.c\in\+W$) is thus $\abs{\+W}T_{\.ci}$.
The probability that every term $t_{\.ci}$ is obtained to $\epsilon/C$ accuracy is therefore $(1-\delta/\abs{\+W})^{\abs{\+W}}\ge 1 - \delta$. In this case, the weighted sum $\sum_{\.c\+W}\alpha_{\.c}t_{\.ci}$ can be computed classically, and satisfies
\eq{
\sum_{\.c\in\+W}\alpha_{\.c} t_{\.ci} &\le \sum_{\.c\in\+W}\alpha_{\.c} \lp \ip{\Psi_{\.c}}{y_i\Phi(\.x_i)} + \epsilon/C\rp  \\
&= \sum_{\.c\in\+W}\alpha_{\.c}\ip{\Psi_{\.c}}{y_i\Phi(\.x_i)} + \frac{\epsilon}{C}\sum_{\.c\in\+W}\alpha_{\.c} \\
&= \sum_{\.c\in\+W}\alpha_{\.c}\ip{\Psi_{\.c}}{y_i\Phi(\.x_i)} + \epsilon
}
and similarly $\sum_{\.c}\alpha_{\.c}t_{\.ci} \ge  \sum_{\.c\in\+W}\alpha_{\.c}\ip{\Psi_{\.c}}{y_i\Phi(\.x_i)} - \epsilon$.
\end{proof}

\section{Proof of Theorem \ref{thm:hybrid-algo-iterations}}\label{app:proof-of-algo}

The analysis of Algorithm \ref{alg:qc-struct-svm} is based on \cite{joachims1999advances,tsochantaridis2005large}, with additional steps and complexity required to bound the errors due to inner product estimation and projection onto the p.s.d. cone.  

\algits*

\begin{lem}\label{lem:all-succeed}
When Algorithm \ref{alg:qc-struct-svm} terminates successfully after at most $t_{\max}$ iterations, the probability that all inner products are estimated to within their required tolerances throughout the duration of the algorithm is at least $1-\delta$.
\end{lem}

\begin{proof}

Each iteration $t$ of the Algorithm involves

\begin{itemize}

\item $O(t^2)$ inner product estimations $IP_{\epsilon_{J},\delta_J}(\Psi_{\.c},\Psi_{\.c'})$, for all pairs $\.c,\.c'\in\+W_t$. The probability of successfully computing all $t^2$ inner products to within error $\epsilon_{J}$ is at least $(1 - \delta_J)^{t^2} = \lp 1 - \frac{\delta}{2t^2 \,t_{\max}}\rp^{t^2} \ge  1 - \frac{\delta}{2\,t_{\max}}$.

\item $m$ inner product estimations $IP_{\epsilon,\delta_\zeta}\lp \sum_{\.c\in\+W^t}\alpha^{(t)}_{\.c}\Psi_{\.c}, y_i\Phi(\.x_i)\rp$, for $i=1,\ldots, m$.  The probability of all estimates lying within error $\epsilon$ is at least $\lp 1 - \frac{\delta}{2m t_{\max}}\rp^m \ge 1 - \frac{\delta}{2_{t_{\max}}}$.

\end{itemize}
Since the algorithm terminates successfully after at most $t_{\max}$ iterations, the probability that all the inner products are estimated to within their required tolerances is
\eq{
\prod_{t=1}^{t_{\max}}\lp 1 - \frac{\delta}{2t_{\max}}\rp^2 &\ge 1 - \delta 
}
where the right hand side follows from Bernoulli's inequality.

\end{proof}

By Lemma \ref{lem:all-succeed} we can analyze Algorithm \ref{alg:qc-struct-svm}, assuming that all the quantum inner product estimations succeed, i.e. each call to $IP_{\epsilon,\delta}(\.x,\.y)$ produces an estimate of $\ip{\.x}{\.y}$ within error $\epsilon$.  In what follows, let $J_{\+W_t}$ be the $\abs{\+W_t}\times\abs{\+W_t}$ matrix with elements $\ip{\Psi_{\.c}}{\Psi_{\.c'}}$ for $\.c,\.c'\in\+W$, let 
$\delta \hat{J}_{\+W_t}\defeq J_{\+W_t} - \hat{J}_{\+W_t} $.

\begin{lem}\label{lem:delta-j}
$\norm{\delta \hat{J}_{\+W_t}}_\sigma \le \norm{\delta \hat{J}_{\+W_t}}_F \le \frac{1}{C t_{\max}}$, where $\norm{\cdot}_\sigma$ is the spectral norm. 
\end{lem}

\begin{proof}
The relation between the spectral and Frobenius norms is elementary. We thus prove the upper-bound on the Frobenius norm. By assumption, all  matrix elements $\tilde{J}_{\.c\.c'}$ satisfy $\abs{\tilde{J}_{\.c\.c'} - \ip{\Psi_{\.c}}{\Psi_{\.c'}}} \le \epsilon_{J} = \frac{1}{C t t_{\max}}$. Thus,
\eq{
\norm{\delta \hat{J}_{\+W_t}}_F &= \norm{J_{\+W_t} - \hat{J}_{\+W_t}}_F\\
&=\norm{ J_{\+W_t} - P_{S^+_{\abs{\+W_t}}}(\tilde{J})}_F\\
&\le \norm{J_{\+W_t} - \tilde{J}_{\+W_t}}_{F} \\
&\le \abs{\+W_t}\epsilon_{J} \\
&\le \epsilon_{J}t \\
&=\frac{1}{C t_{\max}}
}
where the second equality follows from the definition of $\hat{J}_{\+W_t}$ in Algorithm \ref{alg:qc-struct-svm}, the first inequality because projecting $\tilde{J}_{\+W}$ onto the p.s.d cone cannot increase its Frobenius norm distance to a p.s.d matrix $J_{\+W_t}$, and the third inequality because the size of the index set $\+W_t$ increases by at most one per iteration. 
\end{proof}

To proceed, let us introduce some additional notation. Given index set $\+W$, define 
\eq{
D_{\+W}(\alpha) &= -\frac{1}{2}\sum_{\.c, \.c' \in\+W} \alpha_{\.c}\alpha_{\.c'}J_{\.c\.c'} + \sum_{\.c\in\+W} \frac{\norm{\.c}_1}{m}\alpha_{\.c}\\
\hat{D}_{\+W}(\alpha) &= -\frac{1}{2}\sum_{\.c, \.c' \in\+W} \alpha_{\.c}\alpha_{\.c'}\hat{J}_{\.c\.c'} + \sum_{\.c\in\+W} \frac{\norm{\.c}_1}{m}\alpha_{\.c}
}
and let $D_{\+W}^*$ and $\hat{D}_{\+W}^*$ be the maximum values of $D_{\+W}(\alpha)$ and $\hat{D}_{\+W}(\alpha)$ respectively, subject to the constraints $\alpha \ge 0, \sum_{\.c\in\+W}\alpha_{\.c} \le C$. Since $\hat{J}$ above is positive semi-definite, its matrix elements can expressed as 
\eql{
\hat{J}_{\.c\.c'} &= \ip{\hat{\Psi}_{\.c}}{\hat{\Psi}_{\.c'}} \label{eq:psi-hat}
} 
for some set of vectors $\{\hat{\Psi}_{\.c}\}$.  


The next lemma shows that the solution $\hat{\alpha}^{(t)}$ obtained at each step is only slightly suboptimal as a solution for the restricted problem $D_{\+W_t}$.

\begin{lem} \label{lem:opt-approx} $D^*_{\+W_t} - \frac{C}{t_{\max}}\le D_{\+W_t}(\hat{\alpha}^{(t)}) \le D^*_{\+W_t}$. 
\end{lem}

\begin{proof} 
\eql{
D_{\+W_t}^* &\ge D_{\+W_t}\lp \hat{\alpha}^{(t)}\rp \nonumber \\ 
&= -\frac{1}{2} \lp\hat{\alpha}^{(t)}\rp^T\lp \hat{J}_{\+W_t} + \delta \hat{J}_{\+W_t} \rp\hat{\alpha}^{(t)} + \sum_{\.c\in\+W}\frac{\norm{\.c}_1}{m}\hat{\alpha}^{(t)}_{\.c} \nonumber \\
&= \hat{D}^{*}_{\+W_t} - \frac{1}{2} \lp\hat{\alpha}^{(t)}\rp^T\lp \delta \hat{J}_{\+W_t}\rp \hat{\alpha}^{(t)} \nonumber \\
& \ge \hat{D}^{*}_{\+W_t} - \frac{1}{2}\norm{\delta \hat{J}_{\+W_t}}_\sigma\norm{\hat{\alpha}^{(t)}}_2^2 \nonumber \\
&\ge \hat{D}^{*}_{\+W_t} - \frac{C^2}{2}\norm{\delta \hat{J}_{\+W_t}}_\sigma \label{eq:f-opt}
}
The first inequality follows from the fact that $\hat{\alpha}^{(t)}$ is, by definition, optimal for $\hat{D}_{\+W_t}$ and feasible for $D_{\+W_t}$, and the last inequality comes from the fact that $\norm{\hat{\alpha}^{(t)}}_2 \le \norm{\hat{\alpha}^{(t)}}_1 \le C$. Similarly, 
\eql{
\hat{D}_{\+W_t}^* &\ge D_{\+W_t}^* - \frac{C^2}{2}\norm{\delta \hat{J}_{\+W_t}}_\sigma \label{eq:f-hat-opt}
}
and the result follows from substituting \eqref{eq:f-hat-opt} into \eqref{eq:f-opt}, and using lemma \ref{lem:delta-j}.
\end{proof}

We now show that $\hat{\alpha}^{(t)}$ and $\hat{\xi}^{(t)}$ can be used to define a feasible solution for OP\ref{op:struct-svm-p} where the constraints are restricted to only hold over the index set $\+W_t$.

\begin{lem}\label{lem:grad-c-in-w} Define $\.w^{(t)} \defeq \sum_{\.c\in\+W_t}\hat{\alpha}_{\.c}^{(t)}\Psi_{\.c}$. It holds that $\frac{1}{m}\sum_{i=1}^m c_i - \ip{\.w}{\Psi_{\.c}} \le \hat{\xi}^{(t)}$ for all $\.c\in\+W_t$.
\end{lem}

\begin{proof}

First note that
\eql{
\sum_{\.c'\in\+W_t}\hat{\alpha}^{(t)}_{\.c'}\lp \delta \hat{J}_{\+W_t}\rp_{\.c^*\.c'} &= \ip{\lp \delta \hat{J}_{\+W_t}\rp_{\.c*}}{\hat{\alpha}^{(t)}} \nonumber\\
&\ge - \norm{\lp\delta\hat{J}_{\+W_t}\rp_{\.c*}}_2\norm{\hat{\alpha}^{(t)}}_2 \nonumber\\
&\ge - C\norm{\lp\delta\hat{J}_{\+W_t}\rp_{\.c*}}_2\nonumber\\
&\ge - C\norm{\delta\hat{J}_{\+W_t}}_F\nonumber\\
&\ge -\frac{1}{t_{\max}} \label{eq:t_t_max_lb}
}
where the second inequality is due to $\norm{\hat{\alpha}^{(t)}}_2 \le \norm{\hat{\alpha}^{(t)}}_1 \le C$, the third is because $\norm{\lp\delta\hat{J}_{\+W_t}\rp_{\.c*}}_2 \le \norm{\delta\hat{J}_{\+W_t}}_F$, the fourth follows from Lemma \ref{lem:delta-j}.

Let $\.c^* = \arg\max_{\.c\in\+W_t}\lp\frac{1}{m}\sum_{i=1}^m c_i  - \sum_{\.c'\in\+W}\hat{\alpha}^{(t)}_{\.c'}J_{\.c\.c'}\rp$. Then,
\eq{
\hat{\xi}^{(t)}&\defeq \max_{\.c\in\+W_t} \lp \frac{1}{m}\sum_{i=1}^m c_i - \sum_{\.c'\in\+W}\hat{\alpha}^{(t)}_{\.c'}\lp\hat{J}_{\+W_t}\rp_{\.c\.c'}\rp +  \frac{1}{t_{\max}}\\
&\ge \frac{1}{m}\sum_{i=1}^m c^*_i - \sum_{\.c'\in\+W_t}\hat{\alpha}^{(t)}_{\.c'}\lp\hat{J}_{\+W_t}\rp_{\.c^*\.c'}+ \frac{1}{t_{\max}} \\
&=\frac{1}{m}\sum_{i=1}^m c^*_i - \sum_{\.c'\in\+W_t}\hat{\alpha}^{(t)}_{\.c'}\lp J_{\+W_t} -\delta \hat{J}_{\+W_t}\rp_{\.c^*\.c'}+ \frac{1}{t_{\max}} \\
&=\max_{\.c\in\+W_t}\lp\frac{1}{m}\sum_{i=1}^m c_i  - \sum_{\.c'\in\+W_t}\hat{\alpha}^{(t)}_{\.c'}J_{\.c\.c'}\rp + \sum_{\.c'\in\+W_t}\hat{\alpha}^{(t)}_{\.c'}\lp \delta \hat{J}_{\+W_t}\rp_{\.c^*\.c'}+ \frac{1}{t_{\max}}\\
&\ge \max_{\.c\in\+W_t}\lp\frac{1}{m}\sum_{i=1}^m c_i  - \sum_{\.c'\in\+W_t}\hat{\alpha}^{(t)}_{\.c'}J_{\.c\.c'}\rp\\
&= \max_{\.c\in\+W_t} \lp \frac{1}{m}\sum_{i=1}^m c_i - \ip{\.w}{\Psi_{\.c}}\rp
}
where the last inequality follows from \eqref{eq:t_t_max_lb}.
\end{proof}

The next lemma shows that at each step which does not terminate the algorithm, the solution $(\hat{\.w}^{(t)}\defeq \sum_{\.c}\hat{\alpha}^{(t)}_{\.c}\Psi_{\.c}, \hat{\xi}^{(t)})$ violates the constraint indexed by $\.c^{(t+1)}$ in OP \ref{op:struct-svm-p} by at least $\epsilon$.

\begin{lem}\label{lem:violate-primal-constraint} 
\eq{
\frac{1}{m}\sum_{i=1}^m\max \left\{0, 1 - \zeta_i\right\} > \hat{\xi}^{(t)} + 2\epsilon &\Rightarrow \frac{1}{m}\sum_{i=1}^n c^{(t+1)}_i -\ip{\.w^{(t)}}{\Psi_{\.c^{(t+1)}}} > \hat{\xi}^{(t)} + \epsilon,
} 
where $\hat{\.w}^{(t)}\defeq \sum_{\.c}\hat{\alpha}^{(t)}_{\.c}\Psi_{\.c}$. 
\end{lem}

\begin{proof}
Algorithm \eqref{alg:qc-struct-svm} assigns the values
\eq{
\zeta_i &\leftarrow IP_{\epsilon, \delta_\zeta}\lp \sum_{\.c\in\+W^t}\hat{\alpha}^{(t)}_{\.c}\Psi_{\.c}, y_i\Phi(\.x_i)\rp \\
c_i^{(t+1)}&\leftarrow \begin{cases} 1 & \zeta_i < 1 \\ 0 & \text{otherwise}\end{cases}
}
Assuming $\hat{\xi}^{(t)} + 2\epsilon  < \frac{1}{m}\sum_{i=1}^m\max \left\{0, 1 - \zeta_i\right\}$, it follows that
\eq{
\hat{\xi}^{(t)} + 2\epsilon  &< \frac{1}{m}\sum_{i=1}^m\max \left\{0, 1 - \zeta_i\right\} \\
&=\frac{1}{m}\sum_{i=1}^m c^{(t+1)}_i(1 - \zeta_i)\\
&=\frac{1}{m}\sum_{i=1}^m c_i^{(t+1)} -  \frac{1}{m}\sum_{i=1}^m c_i^{(t+1)} IP_{\epsilon, \delta_{\zeta}}\lp \sum_{\.c\in\+W^t}\hat{\alpha}^{(t)}_{\.c}\Psi_{\.c}, y_i\Phi(\.x_i)\rp\\
&\le \frac{1}{m}\sum_{i=1}^m c_i^{(t+1)}  - \frac{1}{m}\sum_{i=1}^m c_i^{(t+1)} \lp\ip{ \sum_{\.c\in\+W^t}\hat{\alpha}^{(t)}_{\.c}\Psi_{\.c}}{y_i\Phi(\.x_i)} -\epsilon\rp\\
&= \frac{1}{m}\sum_{i=1}^m c_i^{(t+1)} - \ip{\sum_{\.c\in\+W_t}\hat{\alpha}^{(t)}_{\.c}\Psi_{\.c}}{\frac{1}{m}\sum_{i=1}^m c_i^{(t+1)} y_i \Phi(\.x_i)}+ \epsilon\\
&= \frac{1}{m}\sum_{i=1}^m c_i^{(t+1)} - \ip{\.w^{(t)}}{\Psi_{\.c^{(t+1)}}} + \epsilon
} 
\end{proof}

Next we show that each iteration of the algorithm increases the working set $\+W$ such that the optimal solution of the restricted problem $D_{\+W}$ increases by a certain amount. Note that we do not explicitly compute $D^*_{\+W}$, as it will be sufficient to know that its value increases each iteration.
\begin{lem} \label{lem:min-increase}
While $\xi^{(t)} > \hat{\xi} + \epsilon + \epsilon_c$, $D^*_{\+W_{t+1}} - D^*_{\+W_t} \ge \min\left\{\frac{C\epsilon}{2}, \frac{\epsilon^2}{8R^2}\right\}-\frac{C}{t_{\max}}$
\end{lem}

\begin{proof} 
Given $\hat{\alpha}^{(t)}$ at iteration $t$, define $\alpha,\eta \in\mb{R}^{2^m}$ by
$$\alpha_{\.c} =
\begin{cases}
\hat{\alpha}_{\.c} & \qquad \.c \in \+W_t \\
0 & \qquad \text{o.w.}
\end{cases} \quad\quad\quad\quad
\eta_{\.c} =
\begin{cases} 1 &\qquad \.c= \.c^{(t+1)} \\
-\frac{\alpha_c}{C} & \qquad \.c \in \+W_t \\
0 & \qquad \text{o.w.}
\end{cases}$$

For any $0 \le \beta\le C$, the vector $\alpha + \beta \eta $ is entrywise non-negative by construction, and satisfies 
\eq{
\sum_{\.c\in\{0,1\}^m}\lp \alpha + \beta \eta\rp_{\.c} &= \beta + \lp 1- \frac{\beta}{C}\rp\sum_{\.c\in \+W_t}\hat{\alpha}_{\.c}\\
& \le \beta + C\lp 1- \frac{\beta}{C}\rp \\
&= C
}
$\alpha + \beta \eta$ is therefore a feasible solution of OP \ref{op:struct-svm-d}. Furthermore, by considering the Taylor expansion of the OP \ref{op:struct-svm-d} objective function $D(\alpha)$ it is straightforward to show that
\eql{
\max_{0 \le \beta \le C}  \lp D (\alpha + \beta\eta) - D(\alpha)\rp \ge \frac{1}{2}\min\left\{C, \frac{\eta^T \nabla D(\alpha)}{\eta^T J \eta}\right\}\eta^T\nabla D(\alpha) \label{eq:grad-des}
}
for any $\eta$ satisfying $\eta^T\nabla D(\alpha) > 0$ (See Appendix \ref{app:proof-displacement}). We now show that this condition holds for the $\eta$ defined above. The gradient of $D$ satisfies

\eq{
\nabla D(\alpha)_{\.c} &= \frac{1}{m}\sum_{i=1}^m c_i - \sum_{\.c'\in \+W_t}\alpha_{\.c'}J_{\.c\.c'} \\
&= \frac{1}{m}\sum_{i=1}^m c_i - \ip{\.w^{(t)}}{\Psi_{\.c}}
}

From Lemmas \ref{lem:grad-c-in-w} and \ref{lem:violate-primal-constraint} we have
$$\nabla D(\alpha)_{\.c}  \begin{cases} \le\hat{\xi}^{(t)} &\qquad \.c\in\+W_t\\
\ > \hat{\xi}^{(t)} + \epsilon &\qquad \.c = \.c^{(t+1)}
\end{cases}$$

and since $\sum_{\.c\in\+W_t}\alpha_{\.c} = \sum_{\.c\in\+W_t}\hat{\alpha}_{\.c}\le C$ it follows that  
\eql{
\eta^T\nabla D(\alpha) &= \lp \hat{\xi}^{(t)} + \epsilon\rp - \frac{1}{C}\sum_{\.c\in\+W_t}\alpha_{\.c} \nabla D(\alpha)_{\.c}  \nonumber \\
&\ge \lp \hat{\xi}^{(t)} + \epsilon\rp - \frac{ \hat{\xi}^{(t)} }{C}\sum_{c\in\+W_t}\alpha_{\.c} \nonumber\\
&= \epsilon\label{eq:grad-f}
}

Also:

\eql{
\eta^T J \eta &= J_{\.c^{(t+1)}\.c^{(t+1)}} + \frac{1}{C^2}\sum_{\.c,\.c'\in \+W_t}\alpha_{\.c} \alpha_{\.c'}J_{\.c\.c'} - \frac{2}{C}\sum_{\.c\in\+W_t}\alpha_{\.c} J_{\.c\.c^{(t+1)}} \nonumber \\
&\le R^2 + \frac{R}{C^2}\sum_{\.c,\.c'\in\+W_t}\alpha_{\.c} \alpha_{\.c'}+ \frac{2R^2}{C}\sum_{\.c\in\+W_t}\alpha_{\.c}\nonumber \\
&\le 4 R^2 &\label{eq:denom}
}
where we note that $J_{\.c\.c'} = \ip{\Psi_{\.c}}{\Psi_{\.c'}} \le \max_{\.c\in\{0,1\}^m}\norm{\Psi_{\.c}}^2\le R^2$. Combining \eqref{eq:grad-des}, \eqref{eq:grad-f}, \eqref{eq:denom} gives 
\eq{
\max_{\beta\in[0,C]}D(\alpha + \beta \eta) - D(\alpha) &\ge \min\left\{\frac{C\epsilon}{2}, \frac{\epsilon^2}{8R^2}\right\} 
}
By construction $\max_{\beta\in[0,C]}D(\alpha + \beta \eta) \le D^{*}_{\+W_{t+1}}$ and Lemma \ref{lem:opt-approx} gives  $D^*_{\+W_t} - \frac{C}{t_{\max}}\le D_{\+W_t}(\hat{\alpha})$. Thus

\eq{
D^*_{\+W_{t+1}} &\ge D(\alpha) +  \min\left\{\frac{C\epsilon}{2}, \frac{\epsilon^2}{8R^2}\right\}\\
&= D_{\+W_t}(\hat{\alpha})+   \min\left\{\frac{C\epsilon}{2}, \frac{\epsilon^2}{8R^2}\right\}\\
&\ge D^*_{\+W_t} +  \min\left\{\frac{C\epsilon}{2}, \frac{\epsilon^2}{8R^2}\right\}-\frac{C}{t_{\max}}
}

\end{proof}

\begin{cor}\label{cor:termination}If $t_{\max}\ge  \min\left\{\frac{4}{\epsilon}, \frac{16CR^2}{\epsilon^2}\right\}$, Algorithm \ref{alg:qc-struct-svm} terminates after at most $ t_{\max}$ iterations.
\end{cor}

\begin{proof}
Lemma \ref{lem:min-increase} shows that the optimal dual objective value $D^*_{\+W_t}$ increases by at least $\min\left\{\frac{C\epsilon}{2}, \frac{\epsilon^2}{8R^2}\right\}-\frac{C}{t_{\max}}$ each iteration. For $t_{\max}\ge  \min\left\{\frac{4}{\epsilon}, \frac{16CR^2}{\epsilon^2}\right\}$, this increase is at least $\frac{C}{t_{\max}}$. $D_{\+W_t}^*$ is upperbounded by $D^*$, the optimal value of OP \ref{op:struct-svm-d} which, by Lagrange duality, is equal to the optimum value of the primal problem OP \ref{op:struct-svm-p}, which is itself upper bounded by $C$ (corresponding to feasible solution $\.w = 0, \xi = 1)$.  Thus, the algorithm must terminate after at most $t_{\max}$ iterations.
\end{proof}

We now show that the outputs $\hat{\alpha}$ and $\hat{\xi}$ of Algorithm \ref{alg:qc-struct-svm} can be used to define a feasible solution to OP \ref{op:struct-svm-p}.

\begin{lem}\label{lem:feas} Let $(\hat{\alpha}), \hat{\xi}$ be the outputs of Algorithm \ref{alg:qc-struct-svm} , in the event that the algorithm terminates within $t_{\max}$ iterations. Let $\hat{\.w} = \sum_{\.c}\hat{\alpha}_{\.c}\Psi_{\.c}$. Then $(\hat{\.w}, \hat{\xi} + 3\epsilon)$ is feasible for OP \ref{op:struct-svm-p}.
\end{lem}

\begin{proof} By construction $\hat{\xi} + 3\epsilon >0$. The termination condition $\frac{1}{m}\sum_{i=1}^m \max\left\{0, 1- \xi_i\right\} \le \hat{\xi} + 2\epsilon$ implies that
\eq{
\max_{\.c\in\{0,1\}^m} \lp \frac{1}{m}\sum_{i=1}^m c_i - \ip{\hat{\.w}}{\Psi_{\.c}}\rp&=
\max_{\.c\in\{0,1\}^m} \lp \frac{1}{m}\sum_{i=1}^m c_i - \frac{1}{m}\sum_{i=1}^m c_i y_i \ip{\hat{\.w}}{\Phi(\.x_i)}\rp \\ &=\frac{1}{m}\sum_{i=1}^m\max_{c_i\in\{0,1\}}\lp c_i - c_i\ip{\hat{\.w}}{y_i \Phi(\.x_i)}\rp \\
&\le \frac{1}{m}\sum_{i=1}^m\max_{c_i\in\{0,1\}}c_i\lp 1 -\xi_i + \epsilon \rp\\
&\le \frac{1}{m}\sum_{i=1}^m\max_{c_i\in\{0,1\}}c_i\lp 1 -\xi_i \rp +  \epsilon \\
&= \frac{1}{m}\sum_{i=1}^m \max\left\{0, 1- \xi_i\right\} + \epsilon\\
&\le \hat{\xi} + 3\epsilon
}
$(\hat{\.w}, \hat{\xi} + 3\epsilon)$ therefore satisfy all the constraints of OP \ref{op:struct-svm-p}.
\end{proof}

We are now in a position to prove Theorem \ref{thm:hybrid-algo-iterations}.

\algits*

\begin{proof}
The guarantee of termination with $t_{\max}$ iterations for $t_{\max}\ge \max\left\{\frac{4}{\epsilon}, \frac{16CR^2}{\epsilon^2}\right\}$ is given by Corollary \ref{cor:termination}, and the feasibility of $(\hat{\.w}, \hat{\xi} + 3\epsilon)$ is given by Lemma \ref{lem:feas}.

Let the algorithm terminate at iteration $t \le t_{\max}$. Then, $\hat{D}_{\+W_t}^* = \hat{D}_{\+W_t}\lp \hat{\alpha}^{(t)}\rp= \hat{D}_{\+W_t}\lp \hat{\alpha}\rp$ and, by strong duality, $(\sum_{\.c}\hat{\alpha}_{\.c}\hat{\Psi}_{\.c}, \max_{\.c\in\+W_t}\xi_{\.c}^{(t)})$ is optimal for the corresponding primal problem  
\begin{op}\label{op:struct-svm-p-hat}
\eq{
    \underset{\.w,\ \xi\ge 0}{\min} & \quad  P(\.w, \xi) = \frac{1}{2}\ip{\.w}{\.w} + C\xi \\
    \text{s.t.} &\quad  \frac{1}{m}\sum_{i=1}^m c_i - \ip{\.w}{\hat{\Psi}_{\.c}} \le \xi,\qquad \forall \.c\in\{0,1\}^m.
    }
\end{op}
for $\hat{\Psi}_{\.c}$ defined by \eqref{eq:psi-hat}, i.e.
\eq{
\hat{D}_{\+W_{t}}(\hat{\alpha}) &= \frac{1}{2}\ip{\sum_{\.c}\hat{\alpha}_{\.c}\hat{\Psi}_{\.c}}{\sum_{\.c}\hat{\alpha}_{\.c'}\hat{\Psi}_{\.c'}} + C\max_{\.c\in\+W_t}\xi_{\.c}^{(t)}\\
&= \frac{1}{2}\sum_{\.c\.c'} \hat{\alpha}_{\.c}\hat{\alpha}_{\.c'}\hat{J}_{\.c\.c'} + C\max_{\.c\in\+W_t}\xi_{\.c}^{(t)}
}
Separately, note that
\eql{
\hat{D}_{\+W_t}(\hat{\alpha}^{(t)}) - D_{\+W_t}(\hat{\alpha}^{(t)}) &= \frac{1}{2}\lp\hat{\alpha}^{(t)}\rp^T\lp J_{\+W_t} - \hat{J}_{\+W_t}\rp\hat{\alpha}^{(t)}\nonumber\\
&=-\frac{1}{2}\lp\hat{\alpha}^{(t)}\rp^T\delta \hat{J}_{\+W_t}\hat{\alpha}^{(t)}\nonumber\\
&\le \frac{C^2}{2}\norm{\delta \hat{J}_{\+W_t}}_\sigma\label{eq:alpha-hat}
}
Denote by $\bar{\alpha}$ the vector in $\mb{R}^{2^m}$ given by 
\eq{
\bar{\alpha}_{\.c} &= \begin{cases}\hat{\alpha}_{\.c} & \.c\in\+W_t \\ 0 & \text{otherwise}\end{cases}
}
Then, 
\eq{
P(\hat{\.w}, \hat{\xi}) - P(\.w^*,\xi^*) &= P(\hat{\.w},\hat{\xi}) - D(\alpha^*)\\
&\le P(\hat{\.w},\hat{\xi}) - D(\bar{\alpha}) \\
&= P(\hat{\.w},\hat{\xi}) - D_{\+W_{t}}(\hat{\alpha}^{(t)})\\
&\le P(\hat{\.w},\hat{\xi}) - \hat{D}_{\+W_{t}}(\hat{\alpha}^{(t)}) + \frac{C^2}{2}\norm{\delta \hat{J}_{\+W_{t}}}_\sigma\\
&= \frac{1}{2}\sum_{\.c\.c'}\hat{\alpha}_{\.c}\hat{\alpha}_{\.c'}\lp \Psi_{\.c}^T\Psi_{\.c'}  -\hat{\Psi}_{\.c}^T\hat{\Psi}_{\.c'}\rp + C\lp\hat{\xi} -  \max_{\.c\in\+W_t}\xi_{\.c}^{(t)}\rp+ \frac{C^2}{2}\norm{\delta \hat{J}_{\+W_{t}}}_\sigma\\
&\le  \frac{C}{t_{\max}} + C^2\norm{\delta \hat{J}_{\+W_{t_{\max}}}}_\sigma \\
&\le \frac{2C}{t_{\max}}\\
&= 2\min\left\{\frac{C\epsilon}{4},\frac{\epsilon^2}{16R^2}\right\}
}
The first inequality follows from the fact that $\bar{\alpha}$ is feasible for OP \ref{op:struct-svm-d}. The second inequality is due to \eqref{eq:alpha-hat}. The third comes from the definition $\hat{\xi} -  \max_{\.c\in\+W_t}\xi_{\.c}^{(t)} = \frac{1}{t_{\max}}$
and observing that $\frac{1}{2}\sum_{\.c\.c'}\hat{\alpha}_{\.c}\hat{\alpha}_{\.c'}\lp \Psi_{\.c}^T\Psi_{\.c'}  -\hat{\Psi}_{\.c}^T\hat{\Psi}_{\.c'}\rp = -\frac{1}{2}\lp\hat{\alpha}^{(t)}\rp^T\delta \hat{J}_{\+W_{t}}\hat{\alpha}^{(t)}\le \frac{C^2}{2}\norm{\delta \hat{J}_{\+W_t}}_\sigma$, and the fourth inequality follows from Lemma \ref{lem:delta-j}.

\end{proof}

\section{Proof of Theorem \ref{thm:hybrid-algo-time}}\label{app:hybrid-algo-time}

\algtime*

\begin{proof}
The initial storing of data to qRAM take time $\tilde{O}(md)$.  Thereafter, each iteration $t$ involves \begin{itemize}

\item $O(t^2)$ inner product estimations $IP_{\epsilon_{J},\delta_J}(\Psi_{\.c},\Psi_{\.c'})$, for all pairs $\.c,\.c'\in\+W_t$. By Theorem \ref{thm:q-ip-epsilon}, each requires time $\tilde{O}\lp\frac{\log(1/\delta_J)}{\epsilon_{J}}\frac{R^3}{\min\{\norm{\Psi_{\.c},\norm{\Psi_{\.c'}}}\}}T_\Phi\rp$. By design $\epsilon_{J} = \frac{1}{Ctt_{max}}\ge \frac{1}{Ct_{\max}^2}$ and $\delta_J = \frac{\delta}{2t^2\,t_{\max}} \ge \frac{\delta}{2t_{\max}^3}$. The running time to compute all $t^2$ inner products is therefore $\tilde{O}\lp \frac{CR^3}{\Psi_{\min}}\log\lp \frac{2t^3_{\max}}{\delta}\rp t_{\max}^4T_\Phi\rp$.

\item The classical projection of a $t\times t$ matrix onto the p.s.d cone, and a classical optimization subroutine to find $\hat{\alpha}$.  These take time $O(t^3)$ and $O(t^4)$ respectively,  independent of $m$.

\item Storing the $\hat{\alpha}_{\.c}$ for $\.c\in\+W_t$ and the  $\frac{c_i^{(t+1)}y_i}{\sqrt{m}}\norm{\Phi(\.x_i)}$ for $i=1,\ldots, m$ in qRAM, and computing $\eta_{\.c}$ classically. These take time $\tilde{O}\lp t_{\max}\rp$, $\tilde{O}(m)$ and $O(m)$ respectively.

\item $m$ inner product estimations $IP_{\epsilon,\delta_\zeta}\lp \sum_{\.c\in\+W^t}\alpha^{(t)}_{\.c}\Psi_{\.c}, y_i\Phi(\.x_i)\rp$, for $i=1,\ldots, m$.  By Theorem \ref{thm:q-ip-epsilon-2}, each of these can be estimated to accuracy $\epsilon$ with probability at least $1-\frac{\delta}{2m\,t_{\max}}$ in time $\tilde{O}\lp 
\frac{C\abs{\+W_t}}{\epsilon}\log  \lp\frac{2m\abs{\+W_t}\,t_{\max}}{\delta}\rp \frac{R^3}{\min_{\.c\in\+W_t} \norm{\Psi_{\.c}}} T_\Phi\rp $. As $\abs{\+W_t}\le t_{\max}$, it follows that all $m$ inner products can be estimated in time $\tilde{O}\lp 
\frac{CR^3}{\Psi_{\min}}\frac{mt_{\max}}{\epsilon}\log  \lp\frac{1}{\delta}\rp T_\Phi\rp $.

\end{itemize}

The total time per iteration is therefore
 \eq{
 \tilde{O}\lp \frac{CR^3\log(1/\delta)}{\Psi_{\min}}  \lp  t_{\max}^4 + \frac{mt_{\max}}{\epsilon}\rp T_\Phi  \rp
}
and since the algorithm terminates after at most $t_{\max}$ steps, the result follows.

\end{proof}

\section{Proof of Equation \ref{eq:grad-des}}\label{app:proof-displacement}

Let $D(\alpha) = -\frac{1}{2}\alpha^T J \alpha + c^T\alpha$ where $J$ is positive semi-definite.  Here we show that
\eq{
\max_{0 \le \beta \le C}  \lp D (\alpha + \beta\eta) - D(\alpha)\rp \ge \frac{1}{2}\min\left\{C, \frac{\eta^T \nabla D(\alpha)}{\eta^T J \eta}\right\}\eta^T\nabla D(\alpha) 
}
for any $\eta$ satisfying $\eta^T \nabla D(\alpha) > 0$. The change in $D$ under a displacement $\beta\eta$ for some $\beta \ge 0$ satisfies
\eq{
\delta D \defeq D(\alpha + \beta\eta) -  D(\alpha) &= \beta\eta^T\nabla D(\alpha) - \frac{\beta^2}{2}\eta^T J \eta 
}
which is maximized when 
\eq{
\frac{\pd}{\pd \beta}\delta D &= \eta^T\nabla D(\alpha) - \beta \,\eta^T J\eta =0 \\
\Ra \beta^* &= \frac{\eta^T\nabla D(\alpha)}{\eta^T J\eta}  \\
}
If $\beta^* \le C$ then $\delta D= \frac{1}{2}\frac{\lp \eta^T\nabla D(\alpha)\rp^2}{\eta^T J\eta}$. If $\beta^* > C$ then, as $D$ is concave, the best one can do is choose $\beta= C$, which gives
\eq{
\delta D &= C \eta^T\nabla D(\alpha) - \frac{C^2}{2}\eta^T J \eta \\
&\ge \frac{C}{2}\eta^T\nabla D(\alpha) 
}
where the last line follows from $\beta^* > C \Rightarrow \eta^T\nabla D(\alpha) = \beta^* \eta^TJ \eta \ge C \eta^T J\eta$.

\section{Choice of quantum feature map}\label{app:feature-map}
Let $Z_j$ be the Pauli $Z$ operator acting on qubit $j$ in an $N$ qubit system. Define $M = \frac{1}{N}\lp \sum_j Z_j \rp^2$ and let $M = \sum_{\.c, \.c'\in\{0,1\}^N} M_{\.c\.c'}\ketb{\.c}{\.c'}$ be $M$ expressed in the computational basis. Define the vectorized form of $M$ to be $\ket{M}=\sum_{\.c, \.c'\in\{0,1\}^N}M_{\.c\.c'}\ket{\.c}\ket{\.c'}$. Define the states 
\eq{
\ket{W} &\propto \ket{0}\ket{M} - \mu_0 \ket{1}\ket{\.0}\\
\ket{\Psi}&= \frac{\ket{0}\ket{\psi}\ket{\psi} + \ket{1}\ket{\.0}}{\sqrt{2}}
}
where $\mu_0 > 0$, $\ket{\psi}$ is any $N$ qubit state, and $\ket{\.0}$ is the all zero state on $2N$ qubits. It holds that
\eq{
\braket{W}{\Psi} &\propto \braket{M}{\psi}\ket{\psi} - \mu_0 \\
&= \bra{\psi} M \ket{\psi} - \mu_0
}
Thus
\eq{
\braket{W}{\Psi} &\begin{cases}
\ge 0 &  \bra{\psi} M \ket{\psi} \ge \mu_0\\
< 0  & \bra{\psi} M \ket{\psi} < \mu_0
\end{cases}
}


\end{document}